\documentclass[12pt]{article}

\usepackage{newtxtext,newtxmath}

\usepackage{graphicx}

\usepackage[letterpaper,margin=1in]{geometry}

\renewenvironment{abstract}
	{\quotation}
	{\endquotation}

\date{}

\makeatletter
\renewcommand{\fnum@figure}{\textbf{Figure \thefigure}}
\renewcommand{\fnum@table}{\textbf{Table \thetable}}
\makeatother

\usepackage{scicite}

\usepackage{hyperref}
\usepackage{breakurl}

\usepackage{xspace}
\usepackage{aas_macros}
\newcommand{\celerite}{\texttt{Celerite}\xspace}
\newcommand{\pyttv}{\texttt{PyTTV}\xspace}
\newcommand{\pytransit}{\texttt{PyTransit}\xspace}
\newcommand{\rebound}{\texttt{REBOUND}\xspace}
\newcommand{\reboundx}{\texttt{REBOUNDx}\xspace}
\newcommand{\emcee}{\texttt{emcee}\xspace}
\newcommand{\tess}{TESS\xspace}

\def\scititle{
	Uncovering the Rapidly Evolving Orbits of the Dynamic TOI-201 System
}
\title{\bfseries \boldmath \scititle}

\author{
	Ismael~Mireles$^{1\ast}$,
	Sol\`{e}ne~Ulmer-Moll$^{2}$,
    Donald Liveoak$^{3,4,5}$,
	Diana~Dragomir$^{1}$,\and
    Judith~Korth$^{6}$,
    Alexander~Venner$^{7,8}$,
    Karen~A.~Collins$^{9}$,
    Amaury~H.M.J.~Triaud$^{10}$,\and
    Tristan~Guillot$^{11}$,
    Antoine~Petit$^{11}$,
    Theron~Carmichael$^{12}$,
    Sarah~Millholland$^{3,4}$,\and
    Tim~Hallatt$^{3,4}$,
    Hannu~Parviainen$^{13,14}$,
    Hugh~P.~Osborn$^{15,16}$,
    David~Rapetti$^{17,18}$,\and
    Thomas~A.~Baycroft$^{10,19}$,
    Siddharth~Bhatnagar$^{6,20}$,
    François~Bouchy$^{6}$,\and
    Radka~Dancikova$^{21}$,
    Pedro~Figueira$^{6,22}$, 
    Monika~Lendl$^{6}$,
    St\'ephane~Udry$^{6}$,\and
    Peter~Wheatley$^{23}$,
    Lyu~Abe$^{11}$,
    Abdelkrim Agabi$^{11}$,
    Matteo Beltrame$^{24}$,\and
    Philippe Bendjoya$^{11}$,
    Vincent Deloupy$^{24}$,
    Djamel~M\'{e}karnia$^{11}$,\and
    Fran\c{c}ois-Xavier~Schmider$^{11}$,
    Olga~Su\'{a}rez$^{11}$,
    Khalid~Barkaoui$^{14,25,26}$,
    Keith~Horne$^{27}$,\and
    Felipe~Murgas$^{13,14}$,
    Enric~Palle$^{13,14}$,
    Richard~P.~Schwarz$^{8}$,
    Ramotholo~Sefako$^{28}$,\and
    Avi~Shporer$^{3}$,
    Gregor~Srdoc$^{29}$,
    Chris~Stockdale$^{30}$,
    Francis~P.~Wilkin$^{31}$,\and
    Joel~D.~Hartman$^{32}$, 
    Lauren~A.~Sgro$^{33}$, 
    Thiam-Guan~Tan$^{34}$, 
    Jon~M.~Jenkins$^{17}$,\and
    Attila B\'odi$^{32}$, 
    David Havell$^{35}$,
    Darren Rivett$^{35}$,
    Ian Transom$^{35}$\\
	\small$^{1}$Department of Physics and Astronomy, University of New Mexico, Albuquerque, NM 87106, USA.\and
	\small$^{2}$Leiden Observatory, Leiden University, Leiden, The Netherlands.\and
    \small$^{3}$Department of Physics and Kavli Institute for Astrophysics and Space Research, Massachusetts Institute\and 
    \small of Technology, Cambridge, MA 02139, USA.\and 
    \small$^{4}$MIT Kavli Institute for Astrophysics and Space Research, Massachusetts Institute of Technology, \and 
    \small Cambridge, MA 02139, USA.\and 
    \small$^{5}$ Department of Physics, University of Michigan, Ann Arbor, MI 48109, USA.\and 
    \small$^{6}$Observatoire astronomique de l'Universit\'e de Gen\`eve, Versoix, 1290, Switzerland.\and
    \small$^{7}$ Centre for Astrophysics, University of Southern Queensland, Toowoomba, QLD 4350, Australia. \and 
    \small$^{8}$Max Planck Institute for Astronomy, 69117 Heidelberg, Germany \and 
    \small$^{9}$ Center for Astrophysics~\textbar~Harvard \& Smithsonian, Cambridge, MA 02138, USA.\and 
    \small$^{10}$ School of Physics \& Astronomy, University of Birmingham, Edgbaston, Birmingham, B15 2TT, UK.\and 
    \small$^{11}$ Observatoire de la C\^{o}te d’Azur, UniCA, Laboratoire Lagrange, CNRS UMR 7293, Nice Cedex 4, France.\and 
    \small$^{12}$ Institute for Astronomy, University of Hawai’i, Honolulu, HI 96822, USA.\and 
    \small$^{13}$Departamento de Astrof\'isica, Universidad de La Laguna (ULL), La Laguna, E-38206, Spain.\and 
    \small$^{14}$Instituto de Astrof\'isica de Canarias (IAC), La Laguna, E-38200, Spain.\and 
    \small$^{15}$NCCR/Planet-S, Physikalisches Institut, Universit\"at Bern, Bern, Switzerland.\and 
    \small$^{16}$Institut f\"ur Teilchen- und Astrophysik, ETH Z\"urich, Z\"urich, Switzerland.\and 
    \small$^{17}$NASA Ames Research Center, Moffett Field, CA 94035, USA.\and 
    \small$^{18}$Research Institute for Advanced Computer Science, Universities Space Research Association,\and
    \small Washington, DC 20024, USA.\\ 
    \small$^{19}$Tsung-Dao Lee Institute, Shanghai Jiao Tong University, 1 Lisuo Road, Shanghai 201210, China\and 
    \small$^{20}$Group of Applied Physics and Institute for Environmental Sciences, Universit\'{e} de Gen\`{e}ve, Gen\`{e}ve, Switzerland\and
    \small$^{21}$Institute of Physics, \'Ecole Polytechnique F\'ed\'erale de Lausanne (EPFL), Observatoire de Sauverny,\and
    \small Versoix, Switzerland\\
    \small$^{22}$Instituto de Astrof\'{i}sica de Andaluc\'{i}a-CSIC, Glorieta de la Astronom\'{i}a s/n, E-18008 Granada, Spain\and
    \small$^{23}$Department of Physics, University of Warwick, Coventry CV4 7AL, UK.\and
    \small$^{24}$PNRA \& IPEV, Concordia Station, Antarctica.\and 
    \small$^{25}$Astrobiology Research Unit, Universit\'e de Li\`ege, Li\`ege, Belgium.\and 
    \small$^{26}$Department of Earth, Atmospheric and Planetary Science, Massachusetts Institute of Technology,\and
    \small Cambridge, MA 02139, USA.\and 
    \small$^{27}$SUPA Physics and Astronomy, University of St. Andrews, Fife, KY16 9SS Scotland, UK.\and 
    \small$^{28}$South African Astronomical Observatory, P.O. Box 9, Observatory, Cape Town 7935, South Africa.\and 
    \small$^{29}$Kotizarovci Observatory, Viskovo, Croatia.\\ 
    \small$^{30}$Hazelwood Observatory, Australia.\and 
    \small$^{31}$Department of Physics and Astronomy, Union College, Schenectady, NY 12308, USA.\and 
    \small$^{32}$ Department of Astrophysical Sciences, Princeton University, Princeton, NJ 08544, USA.\and 
    \small$^{33}$SETI Institute, Carl Sagan Center, Mountain View, CA 94043, USA.\and 
    \small$^{34}$Perth Exoplanet Survey Telescope, Perth, Western Australia, Australia.\and 
    \small$^{35}$SETI Institute \& Unistellar Citizen Science Network.\\
	\small$^\ast$Corresponding author. Email: mirelesi@unm.edu\and
}

\begin{document} 

\maketitle


\begin{abstract} \bfseries \boldmath
Studying planetary interactions in exoplanet systems informs theories of planet formation and evolution, providing essential context for understanding our own solar system. We combine spectroscopy, transit photometry, transit timing variations, and astrometry to characterize the TOI-201 system. The cotransiting system consists of a super-Earth, warm Jupiter, and massive companion at 5.8, 53, and 2900 day orbital periods, respectively. We perform dynamical simulations to study the past and future of the system. von-Zeipel-Kozai-Lidov oscillations emerge as the most plausible scenario to explain the outer companion's high orbital eccentricity, with planet-planet scattering a possible but less likely contender. Due to nonzero mutual inclinations between the planets, the system is visibly evolving on very short timescales, with the current cotransiting configuration ending in 200 years.
\end{abstract}

\subsection*{Introduction}

\noindent Most of the giant exoplanets discovered to date have properties that are, for the most part, very different from those of the Solar System gas giants. They tend to orbit much closer to their host star, and are often found in noncircular orbits. When those giant exoplanets reside in multiplanet systems, a wealth of new clues regarding their dynamical evolution becomes available.

The Transiting Exoplanet Survey Satellite, or TESS, has discovered over 650 new planets, 242 of which are part of multiplanet systems. Unlike the previous transit surveys, TESS is an all-sky survey observing stars of all brightness instead of focusing on fainter stars or specific regions of the sky. As a result, TESS has discovered planets around bright stars that are ideal for follow-up observations to characterize the planets in unprecedented detail. One such planet discovered by TESS is TOI-201~b, a warm Jupiter orbiting a relatively bright F-type star at a 53 day period \cite{hobson21}. Warm Jupiters are defined as giant planets with orbital periods between 10 and 200 days. Often described as bridging the gap between hot Jupiters and Jovian analogs, these planets have been subjected to numerous studies over the past decade \cite{dawson2015,huang2016,wu2023}. 

We are beginning to understand how these planets interact with other planets in the same system. Studies of warm Jupiters observed by Kepler have found that the majority of them have small, nearby companions \cite{huang2016, wu2023}. These results suggest that warm Jupiter systems likely formed in situ \cite{boley2016} or further out beyond the ice line before undergoing disk-driven migration \cite{walsh2011, hallatt2020}, as these mechanisms tend to preserve nearby planets. This is in contrast to hot Jupiters, whose general lack of nearby companions point to more dynamically violent mechanisms as the origin of those systems \cite{ford2008origins, petrovich2015hot, vick2019chaotic}. While warm Jupiters are more likely to have nearby small companions than hot Jupiters, the opposite is true for massive, distant companions. Studies have found that hot Jupiters are more likely to have massive (between 1 and 20 Jupiter masses), distant (between 1 and 20 AU) companions than warm Jupiters, although the difference in their occurrence rate is not as substantial as the difference in occurrence rate of small planets \cite{bryan2016}. Nonetheless, this difference again suggests distinct mechanisms at play for hot and warm Jupiter systems, as massive, distant companions drive the more dynamically active pathways \cite{dawson2018}. Finding systems with both nearby and distant companions will provide a more complete picture of warm Jupiter formation pathways. 

Among warm Jupiters with outer companions, there is a small but growing sample of systems in which the companions have masses close to the brown dwarf lower mass limit. These systems open the door to potential secular interactions that could be probed observationally on timescales as short as a human lifetime.

\subsection*{Results}
\subsubsection*{Identifying Two New Companions}

The TESS mission originally identified the warm Jupiter TOI-201~b as a candidate on 2019-05-07 \cite{Guerrero2021}. It was later confirmed as a planet with a mass of 0.42 Jupiter masses and a moderately eccentric orbit ($e\sim0.28$) \cite{hobson21}. TESS later identified a second candidate in the system, TOI-201~d, on 2020-03-11, a potential super-Earth with an orbital period of 5.85 days. While this candidate was known at the time of the confirmation of the warm Jupiter, it was not confirmed, as its radial velocity (RV) signal was too weak to be detected. This is partly due to a linear trend seen in the RVs that was attributed solely to stellar activity. However, as we describe here, the trend was due to an additional massive planet exterior to the warm Jupiter. Here, we statistically validate TOI-201~d using the \texttt{triceratops} package \cite{triceratops1} and obtain a tentative mass measurement.

We visually identified a single, partial transit event in TESS Sector 64 that was unrelated to the super-Earth and warm Jupiter and corresponded to the recently confirmed TOI-201~c \cite{2025arXiv250711504M}. The transit coincided with variations in the timing of the transits of the warm Jupiter, with the transits immediately after the single transit occurring about 30 minutes later than expected. The fact that these sudden transit timing variations (TTVs) occurred so close to the single transit suggested that whatever caused the single transit was also responsible for TOI-201~b's TTVs.

\subsubsection*{Determining the Orbital Period of the Outer Companion}

Having transited only once and partially during TESS observations, we had minimal constraints on the orbital period of the outer companion. However, we used the duration of the transit to obtain a rough estimate of the orbital period. 
Assuming a circular orbit, the 13-hour transit duration implied an orbital period of approximately 250 days \cite{2003ApJ...585.1038S}. To motivate our follow-up observation strategy, we used the TESS data to determine where additional transits could have fallen into data gaps. The shortest possible orbital period was 200 days and there were many possible periods below 500 days that could be observed using ground-based photometry. We observed some from the ground using a variety of telescopes across the world to try to detect a second transit. The telescopes were those from the Las Cumbres Observatory Global Telescope Network (LCOGT), Perth Exoplanet Survey Telescope (PEST), Hazelwood Observatory, HATPI, and Unistellar Network. At the same time, we monitored the system using RV measurements from the CORALIE spectrograph on the Swiss 1.2-m Leonhard Euler Telescope and HARPS spectrograph on the European Southern Observatory’s (ESO) 3.6-m Telescope, both at ESO La Silla Observatory in Chile, as well as Carnegie's Planet Finder Spectrograph (PFS) attached to the 6.5-m Magellan Clay telescope at Las Campanas Observatory in Chile. The photometric observations revealed no new transits while the RVs showed a 200 m/s drop nearly 4 years after the RV observations from the confirmation paper, which indicates that the orbital period was at least a few years. Following a year of RV monitoring, we were finally able to constrain the orbital period of the outer companion to approximately 2900 days.

\subsubsection*{Monitoring TTVs of TOI-201~b}

The single transit of the outer companion coincided with the start of significant TTVs of the warm Jupiter. While early transits occur when predicted, the warm Jupiter's transits immediately before and after the outer companion's single transit varied, with the one before occurring minutes earlier than predicted and the two after occurring later. These variations could not be explained by a slightly erroneous ephemeris or period, indicating that the warm Jupiter and outer companion are dynamically interacting. Unfortunately, the system would stop being observed by TESS soon after, as it did not observe TOI-201 between Sectors 69 and 86. Accordingly, we monitored the warm Jupiter TTVs using ground-based facilities, namely, the LCOGT and the Antarctic Search for Transiting ExoPlanets (ASTEP) telescope located at Concordia station in Antarctica \cite{dransfield2022, schmider2022}. We observed an additional eight transits using LCOGT and ASTEP and eventually obtained an additional transit from TESS when it re-observed the system. These newest observations showed that the TTVs had decreased from their peak immediately after the single transit and appeared to show a gradual decline as the corrected early transits did, as shown in Fig. \ref{fig:ttvs_astrometry}.

\subsubsection*{Modeling the TOI-201 System}

We model the host star's parameters by using archival spectroscopy, photometry, and astrometry to perform an isochrone fit. We find the host star is an F-type star that is slightly larger and hotter than the Sun ($R_\star = 1.31\, R_\odot$, $M_\star = 1.32\, M_\odot$, $T_\star = 6423\, \mathrm{K}$). We also find that the star is relatively young, although the exact age is not well-constrained at $666^{+673}_{-442}$ Myr.

The host star TOI-201 has been found to be a \textit{Hipparcos-Gaia} astrometric accelerator, meaning that its proper motion changed noticeably between observations from the \textit{Hipparcos} and \textit{Gaia} missions \cite{Kervella2019, Brandt2018, Brandt2021}. This astrometric acceleration can be attributed to an unseen massive, distant companion. We modeled the plausible range of mass functions across different orbital separations (as in \cite{Kervella2019}), and, as Fig.~\ref{fig:ttvs_astrometry} shows, the properties of TOI-201~c can wholly reproduce the observed acceleration. As such, we also jointly modeled the \textit{Hipparcos-Gaia} astrometry with the RVs and the single transit from TESS to characterize the orbit of TOI-201~c, allowing us to directly constrain for the outer companion an orbital element inaccessible to the transit and RV methods: the longitude of the ascending node ($\Omega$), the last quantity needed to determine the full three-dimensional orbit of TOI-201~c. We obtained a well-constrained measurement of $\Omega_c$ = 211 $\pm$ 11 deg. Combined with the difference in $\Omega$ between TOI-201~b and c constrained by the photodynamical and RV joint fit, we also obtain $\Omega_b$ = $198\pm10$ deg. To our knowledge, this is the first constraint on the absolute value of $\Omega$ for a warm Jupiter planet.

We fit the aforementioned TESS photometry, RVs from CORALIE, HARPS, and PFS, LCOGT and ASTEP photometry we acquired for TOI-201~b, alongside with archival LCOGT and Next-Generation Transit Survey (NGTS) photometry for TOI-201~b and archival RVs from FEROS and MINERVA-Australis to determine the orbital and physical parameters of the three known bodies orbiting TOI-201. We used the Python Tool for Transit Variations (\texttt{pyTTV}) to perform photodynamical modeling of the photometry jointly with the RVs. The resulting best-fit parameters and associated uncertainties are listed in Table \ref{tab:main_params}, and the data and best-fit models are shown in Fig. \ref{fig:data-and-models}.

We find very similar values for the orbital parameters (period, eccentricity, argument of periastron) of the warm Jupiter as the discovery paper, though we obtain tighter constraints on both the eccentricity and argument of periastron. We obtain a mass for the warm Jupiter of $164\pm5$ Earth masses ($0.52\pm0.02$ Jupiter masses). For the super-Earth, we obtain a relatively weakly constrained mass of $5.8\pm2$ Earth masses. Combined with its radius of $1.39\pm0.07$ Earth radii, we obtain a relatively high bulk density of $11\pm4$ g cm$^{-3}$, twice that of Earth. We also find that its orbit is moderately eccentric ($e =0.3\pm0.1$). The outer companion is the longest-period transiting body found by TESS to date, with an orbital period of $2890\pm20$ days and corresponding semi-major axis of $4.37\pm0.04$ AU. We improve the precision on the period by a factor of 10 compared to the literature \cite{2025arXiv250711504M}. This will be improved even further with observations of its next transit on 2031-03-26. We determine its mass to be $4990\pm100$ Earth masses, or $15.7\pm0.3$ Jupiter masses, placing it just above the deuterium mass burning limit of $\sim13$ Jupiter masses that separates planets from brown dwarfs \cite{2011ApJ...727...57S}. Its orbit is highly eccentric, with an eccentricity of $0.651\pm0.006$, with its closest and furthest approach from the host star bringing it to closer than Mars's orbit and further than Jupiter's, respectively. 

We are also able to place constraints on the mutual inclinations between the different planets, which is the three-dimensional angle between the orbital planes of two planets and is given by the equation

$$\cos \Delta i_{1,2} = \cos i_1\, \cos i_2 + \sin i_1\, \sin i_2\, \cos(\Omega_1-\Omega_2).$$

\noindent We find that the warm Jupiter and brown dwarf have a mutual inclination of $13.4^{+2.0}_{-2.3}$ $^\circ$, the warm Jupiter and super-Earth have a mutual inclination of $28.0^{+10.6}_{-14.2}$ $^\circ$, and the super-Earth and brown dwarf have a mutual inclination of $41.8^{+10.9}_{-16.4}$ $^\circ$. Our mutual inclination between the warm Jupiter and brown dwarf is notionally over 5$\sigma$ from zero, and differs by more than 2$\sigma$ from the value from the literature \cite{2025arXiv250711504M}. However, when we consider a more limited dataset consisting only of the archival RVs and TESS photometry, we obtain a value consistent with the literature result. Hence, we suggest that the difference from the literature result stems from the larger observational datasets obtained for this work, and indicates a moderate but statistically significant mutual inclination between the orbits in the TOI-201 system.

\subsection*{Discussion}
\subsubsection*{Evolutionary history}\label{sec:evo_history}

The elevated orbital eccentricity of the brown dwarf is indicative of a dynamically hot past. Several mechanisms are known to increase orbital eccentricity. We rule out interactions with the disk because they require a cavity interior to the brown dwarf's orbit \cite{romanova2023eccentricity}, which is not allowed given the existence of the two interior planets. A stellar flyby is highly unlikely due to the unrealistically tight distance of closest approach required to generate the eccentricity of TOI-201~c, and we find that high-eccentricity migration would have resulted in the ejection of TOI-201~d. Two plausible scenarios remain: planet-planet scattering \cite{chatterjee2008dynamical, carrera2019planet} and von-Zeipel-Lidov-Kozai (vZLK) cycles \cite{takeda2005high}.

We examined the possibility that the brown dwarf obtained its eccentricity through planet-planet scattering with a now-ejected third giant planet early in the system's history. We carried out a suite of $N$-body simulations, varying the initial orbital properties of the four bodies as well as the mass of the ejected planet. While it is possible to obtain the observed parameters via this scenario, only $\sim$1\% of the simulations reproduce the present-day TOI-201 system. Our methodology and results are described in more detail in Supplementary Text.

The dynamical architecture of the TOI-201 system could be sculpted by an as of yet undetected stellar companion. Such an unseen stellar companion is capable of inciting vZLK oscillations in TOI-201~c  \cite{fabtre07,itoohtkat19}. Such vZLK cycles can not only explain the high eccentricity of TOI-201~c but also those of TOI-201~b and d, as a result of their interaction with the outer, eccentric giant planet.

We investigated the dynamical influence of a stellar companion on the TOI-201 planets using the N-body code \texttt{REBOUND} \cite{rein2012rebound,rein2015whfast}. We selected parameters for our hypothetical stellar companion from the constraints based on analysis of the RVs, high-resolution imaging, and Gaia imaging and astrometry (see Fig. \ref{fig:molusc}). 
Specifically, we adopted a companion mass $M_{\rm comp}{=}0.1 \ M_{\odot}$, initial orbital semi-major axis sampled uniformly from $a_{\rm comp}{\in}[80, 120]$ AU, eccentricity $e_{\rm comp}{=}0$, and inclination uniformly sampled in $i_{\rm comp}{\in}[55,65]^\circ$. Planet semi-major axes were sampled within their uncertainties, inclinations were sampled between $0^\circ$ and $5^\circ$, the eccentricity for TOI-201~d was sampled uniformly between 0 and 0.1, and eccentricities for TOI-201~b and c were sampled uniformly between 0 and 0.2. The nonzero primordial eccentricities we employed can plausibly be excited via interaction with the protoplanetary disk \cite{lilai23} and/or through planet-planet dynamical excitation \cite{jurtre08}, which may be a common feature of systems with multiple giant planets such as TOI-201 \cite{frejanmur19}.

\begin{figure}
\centering
\includegraphics[width=\linewidth]{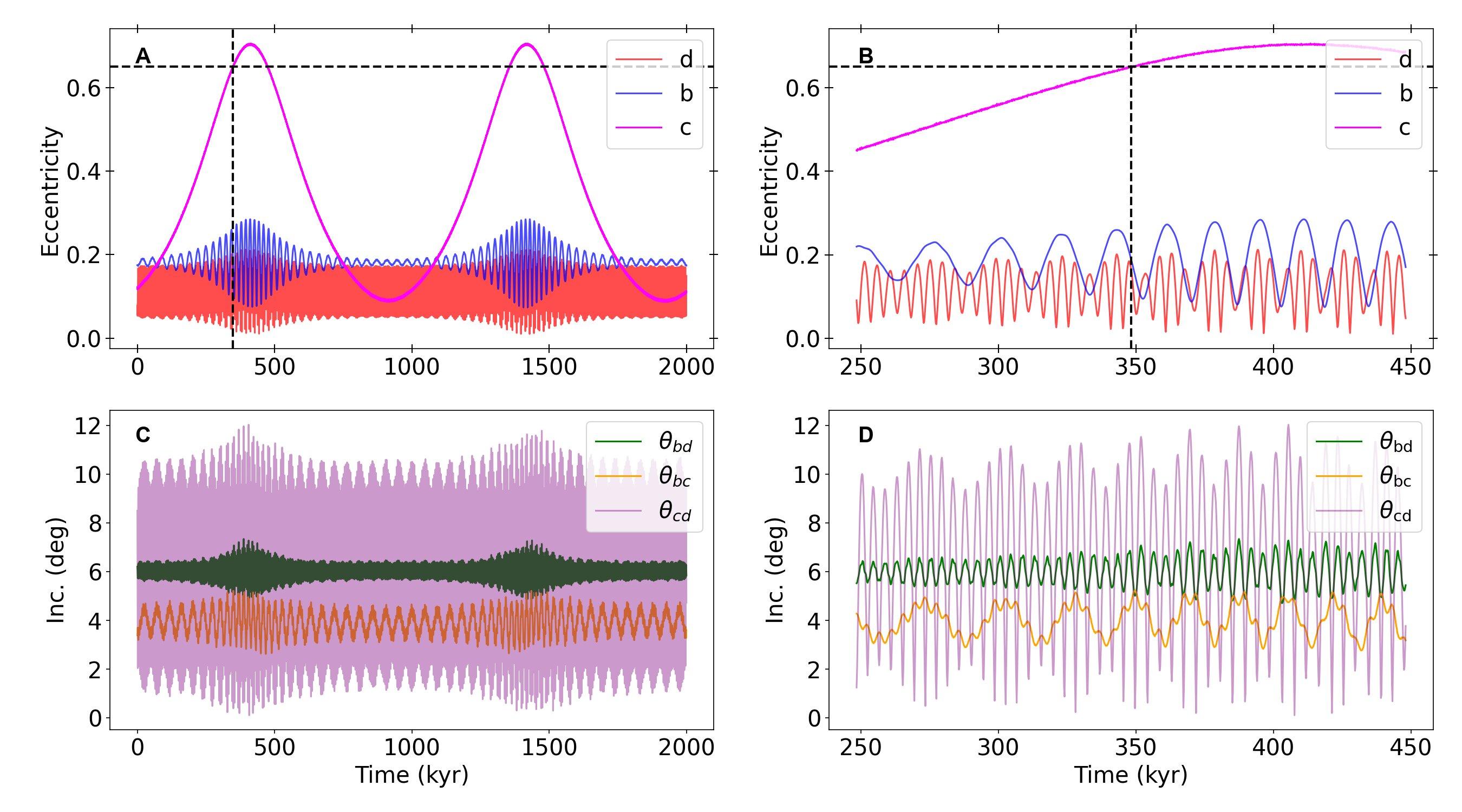}
\caption{\textbf{Dynamical evolution of vZLK simulation which replicates the observed system architectures.} The system contains a hypothetical undetected binary companion of mass $0.1 M_\odot$, semi-major axis $102$ AU, and initial mutual inclination $64^{\circ}$. Dashed lines in (A) and (B) indicate the first time for which $e_\text{c} = 0.65$. (\textbf{A}) The evolution of the eccentricities over the span of 2000 kyr (2 Myr). (\textbf{B}) Zoom-in of (A) between 250 and 450 kyr when the eccentricity of TOI-201 c first reaches 0.65 showing the short timescale evolution. (\textbf{C}) The evolution of the mutual inclinations over the span of 2000 kyr. (\textbf{D}) Zoom-in of (C) between 250 and 450 kyr showing the short timescale evolution.}
\label{fig:ZLK-timeseries}
\end{figure}

Figure \ref{fig:ZLK-timeseries} showcases a representative example of the dynamical evolution of the TOI-201 planets in the presence of an outer stellar companion. As expected under vZLK cycles, the eccentricity and inclination of TOI-201~c undergo periodic oscillations over the vZLK timescale ${\sim}300$ kyr. The observed eccentricity for TOI-201~c, $e_{\rm c}{\sim}0.65$ (see Table \ref{tab:main_params}), is readily attained near the maximum of its vZLK cycle. During its high-eccentricity phases, TOI-201 c pumps the eccentricities and mutual inclinations of the inner planets to their observed values within their respective uncertainties ($e_{\rm b}=0.275\pm0.009$, $e_{\rm d}=0.3\pm0.1$, $\Delta i_{bc}=13{\pm}22$ deg., $\Delta i_{cd}=41{\pm}29$ deg., $\Delta i_{bd}=28{\pm}28$ deg.). Although the resulting mutual inclinations are systematically smaller than the observed values, exploration of a wider range of initial conditions (e.g., an initially misaligned inner system or greater companion inclination) could reveal more extreme misalignment.

The vast majority of our simulated systems remain dynamically stable over Myr timescales, the longest duration of integration we explored. We conclude that vZLK oscillations are the most plausible explanation for TOI-201~c's high eccentricity.

Follow-up observational work could verify the vZLK hypothesis if a stellar companion is found. Measurements of TOI-201's stellar obliquity could also validate the vZLK scenario, since we would expect the orbital angular momentum vector of TOI-201~c to currently be misaligned with the host star's spin axis.

Last, it is possible that a combination of planet-planet scattering and vZLK oscillations could also explain TOI-201's architecture, a possibility which should be explored in future studies of this system.

\subsubsection*{Current State and Immediate Future of the System}

Our goal in this section is to characterize the stability and secular dynamical evolution of the planetary system. We show that while the system is likely stable, there is a non-negligible chance of planet d experiencing instability over Myr timescales. We highlight that TOI-201 exhibits significant secular dynamical evolution over human-observable timescales (${\sim}$decades); long-term observation of TOI-201 may therefore provide an unprecedented glimpse into the active lives of planetary systems in real time.

To explore the current state and future evolution of the TOI-201 system, we constructed a suite of 500 \texttt{REBOUND} N-body calculations \cite{rein2012rebound}. Planet orbital elements were sampled within their uncertainties from the posteriors, and orbital integrations were performed for up to 2 Myr. 

We used the \texttt{MEGNO} chaos indicator from \texttt{REBOUNDx} \cite{tamayo2020reboundx} to determine if the system is likely unstable in the present. We find MEGNO scores consistent with stability (see Fig. \ref{fig:high-e-megno}). System stability is also reflected in our N-body simulations, for which only $1\%$ experienced an instability leading to tidal disruption/ejection of planet d during the high-eccentricity vZLK epochs. There is thus a small but nonzero chance of planet d experiencing dynamical upheaval/destruction over Myr timescales.

As illustrated in Fig. \ref{fig:impact-param-hists}, the transit impact parameters and corresponding transit duration variations (TDVs) of planets b and d evolve significantly over ${\sim}$decades. This can be confirmed observationally after the next periastron passage of the brown dwarf in 2031, when TOI-201~b's impact parameter will increase by more than 3-$\sigma$ from its currently measured value. While planet d's impact parameter is also evolving rapidly, its shallow transits result in a large uncertainty for its impact parameter, and thus it will be over 200 years before it deviates by 3-$\sigma$ from its present value. The significant secular evolution of the planetary transits is driven by the planets' large mutual inclinations. We find that planets b, c, and d will cease to cotransit after just ${\sim}$200 yr, and will only re-establish cotransiting geometry after ${\sim}10$ kyr (see Fig. \ref{fig:impact-dynamics}). Changes in transit geometry are particularly acute for planet b, which exhibits step-like perturbations excited at each periapse passage of TOI-201~c. Our integrations therefore indicate that continued monitoring of the system could witness the evolution of the warm Jupiter's stellar obliquity as it is sculpted by the outer brown dwarf. Follow-up measurements of the stellar obliquity are called for. This can be achieved through observations of the Rossiter-McLaughlin (RM) effect; for TOI-201~b, the expected RM effect amplitude is $\sim$30 m/s, which is well within the capabilities of current spectrographs.

The TOI-201 system further underscores how three-dimensional orbital characterization can shed light on the active lives of planetary systems; without three-dimensional orbit information, we find that the system does not evolve sufficiently quickly that we are able to watch its architecture undergo dynamical sculpting in real time.


\subsection*{Materials and Methods}

\subsubsection*{TESS photometry}\label{sec:tess}

In addition to the 14 initial sectors used in the TOI-201~b discovery paper \cite{hobson21}, we use the most recent 18 sectors of TESS photometry obtained for TOI-201 (TIC 350618622). These 18 sectors include nine new transits for the previously identified warm Jupiter TOI-201~b, for a total of 16 transits observed by TESS. 
The TOI-201 data observed at 2-min cadence and the image data were reduced and analyzed by the Science Processing Operations Center (SPOC) \cite{jenkinsSPOC2016} at NASA Ames Research Center. The TESS Science Office reviewed the vetting information and issued an alert on 7 May 2019 for TOI-201 b and on 11 March 2020 for TOI-201.02 \cite{Guerrero2021}. The signals have been repeatedly recovered with different observations. Combining multiple sectors, the SPOC conducted a transit search of Sectors 1 to 68 on 30 October 2023 with an adaptive, noise-compensating matched filter \cite{2002ApJ...575..493J, 2010SPIE.7740E..0DJ, 2020TPSkdph}, producing Threshold Crossing Events for which an initial limb-darkened transit model was fitted \cite{Li:DVmodelFit2019} and a suite of diagnostic tests were conducted to help make or break the planetary nature of the signals \cite{Twicken:DVdiagnostics2018}. The host star is located within 0.62$\pm$2.49 arcsec of the source of the transit signal for TOI-201 b and within 4.29$\pm$4.56 arcsec for TOI-202.02. The transit signatures were also detected in searches of Full Frame Image data by the Quick Look Pipeline (QLP) at MIT \cite{huang2020a, huang2020b}. 
This candidate was not confirmed alongside the warm Jupiter as a one-planet RV model incorporating Gaussian processes for stellar variability was preferred over a two-planet model \cite{hobson21}. However, we find that the signal attributed to stellar variability is better explained by a long-period planet. We visually identified a single, partial transit-like dip in Sector 64 unrelated to TOI-201~b. We use the pre-search data conditioned simple aperture photometry\cite{smith2012, stumpe2012, stumpe2014} (PDCSAP) light curve for the modeling of the majority of the TESS data. Most of a transit of planet b that occurred in Sector 8 occurred while the instrument was turned off.
The transit egress occurred as observations resumed when temperatures were still changing, resulting in a large ramp feature present in both the simple aperture photometry\cite{twicken:PA2010SPIE,morris:PA2020KDPH} (SAP) and PDCSAP light curves (see Fig. \ref{fig:spoc_lc_corrections}). There is also a downlink gap at the start of the single transit in Sector 64 meaning the ingress was not observed. The PDCSAP flux light curve exhibits a steep slope that can be mistaken for an ingress. This feature is not present in the SAP light curve or the Quick-Look Pipeline \cite{huang2020a, huang2020b} (QLP) light curve, indicating that it is an artifact of the PDC process. 
As such, we use light curves corrected using Cotrending Basis Vectors (CBVs) as described next, to model the partial transit of TOI-201~b in Sector 8 and the single transit of TOI-201~c in Sector 64 (Fig. \ref{fig:spoc_lc_corrections}). We used software from \cite{Rapetti2024} that performs systematic corrections and automatically optimizes parameters for correctors available in this code. Here, we focus on a corrector that is a version of PDC adapted within the CBVCorrector class of Lightkurve \cite{Lightkurve}. This corrector uses the CBV technique that the PDC method of the SPOC pipeline uses. Hereafter, we will refer to this corrector as CBV (for comparison purposes, results from other correctors in the code are also shown in Fig. \ref{fig:spoc_lc_corrections}; for further details on these correctors, see the code references above). Flux fraction and crowding adjustments are applied to the corrected light curves. To automatically select optimal values for a set of parameters of the CBVCorrector, each corrected light curve is evaluated using the Savitzky-Golay combined differential photometric precision (sgCDPP) proxy algorithm discussed in \cite{2011ApJS..197....6G, VanCleve:2016} and implemented in Lightkurve, for various durations (see the top panels in Fig. \ref{fig:spoc_lc_corrections}). For a grid of corrector parameter values, the code calculates the harmonic mean (HM) of these sgCDPPs of various durations and selects the corrected light curve that minimizes the HM.

As the single transit in Sector 64 was not associated with either the super-Earth candidate or the confirmed warm Jupiter, the orbital period of the outer candidate was almost completely unconstrained. However, the vast amount of TESS data meant we could determine the minimum orbital period as well as test narrow windows associated with periods where potential additional transits could have fallen into data gaps. To do this, we used the \texttt{MonoTools} package \cite{osborn2022a, osborn2022b} to determine which orbital periods were permitted by the TESS data (see Fig. \ref{fig:monotools}). The package determines the periods allowed by the photometry and calculates a probability for each based on the geometric transit probability and a prior on the eccentricity needed to match the transit duration based on the eccentricity distribution of known planets.  
For systems with multiple transiting planets, \texttt{MonoTools} automatically used an eccentricity prior derived from transiting Kepler planets \cite{van2015eccentricity}; however this is only valid for compact systems of small planets.
This typically results in the posterior period distribution being more tightly distributed around the circular period estimated from the transit model.

\subsubsection*{Ground-based photometry}

We observed potential TTVs in the transits observed by TESS in Sectors 61, 65, and 68 around the time of the single transit event. To determine whether the TTVs were real and characterize them if they were, we observed 15 transits of TOI-201~b from the ground using a combination of the NGTS survey 0.2\,m telescopes \cite{wheatley18ngts} located at ESO's Paranal Observatory, the LCOGT \cite{Brown:2013} 1.0\,m network nodes at Cerro Tololo Inter-American Observatory in Chile (CTIO), Siding Spring Observatory (SSO) near Coonabarabran, Australia, and South Africa Astronomical Observatory (SAAO) near Sutherland, South Africa, and the ASTEP telescope located at Concordia station in Antarctica \cite{Guillot:2015,Mekarnia:2016}. The LCOGT images were calibrated by the standard LCOGT {\tt BANZAI} pipeline \cite{McCully:2018} and differential photometric data were extracted using {\tt AstroImageJ} \cite{Collins:2017}.

We observed two transit windows of TOI-201~d using LCOGT-CTIO and LCOGT-SAAO. The transit event is generally too shallow to be detected by ground-based telescopes. However, we ruled out nearby eclipsing binaries as potential sources of the detection in the \textit{TESS} data. 

We also attempted to search for additional transit events of the brown dwarf TOI-201~c, before its period was known. We searched the shortest possible periods as determined by our \texttt{MonoTools} analysis. We collected nearly two dozen observations over the time period 2023 November 20 to 2024 December 20 using LCOGT-CTIO, LCOGT-SSO, LCOGT-SAAO, the Perth Exoplanet Survey Telescope (PEST) located near Perth, Australia, Hazelwood Observatory near Churchill, Victoria, Australia, HATPI located at Las Campanas Observatory in the Chilean Andes, and from three Unistellar Network telescopes in Australia and New Zealand \cite{Marchis:2020}. We found no transit-like events in the periods we checked, which were different from the period determined in this work. An observation log of all ground-based lightcurve observations is provided in Table \ref{table:ground_phot}.

\subsubsection*{Spectroscopic observations}

We use archival RV measurements from CORALIE, HARPS, FEROS, and MINERVA-Australis in combination with new observations from CORALIE, HARPS, and PFS to characterize the system.

\subsubsection*{CORALIE and HARPS}

We collected 23 new RV measurements between UT 2024 January 02 and UT 2025 April 13 with the CORALIE spectrograph on the Swiss 1.2-m Leonhard Euler Telescope at the ESO La Silla Observatory in Chile \cite{2001Msngr.105....1Q}. We also obtained 14 new measurements between UT 2024 October 20 and UT 2025 March 30 with the HARPS spectrograph on the ESO 3.6-m Telescope, also at La Silla Observatory \cite{2003Msngr.114...20M}. We also include the 13 and 42 archival RVs used in the TOI-201~b discovery paper from CORALIE and HARPS, respectively. 
\subsubsection*{Planet Finder Spectrograph}

We collected 19 RV measurements of TOI-201 between UT 2023 Dec 20 and UT 2024 March 03 with the Carnegie PFS \cite{Crane:2006, Crane:2008, Crane:2010}. PFS is a high-precision echelle spectrograph attached to the 6.5-m Magellan Clay telescope at Las Campanas Observatory in Chile. It has a spectral resolution of 130,000 and covers the 390- to 734-nm spectral window. Wavelength calibration is carried out using an iodine absorption cell, which also allows for characterization of the instrumental profile. Spectra were reduced using the standard PFS reduction pipeline \cite{Butler:1996,Crane:2006} and RV measurements were extracted using a custom IDL pipeline.

\subsubsection*{Archival FEROS and MINERVA-Australis RVs}

Our analysis also includes 52 archival RVs from the Fiber-fed Extended Range Optical Spectrograph (FEROS) at the MPG/ESO 2.2-m telescope at La Silla Observatory \cite{1999Msngr..95....8K} and 62 from the MINERVA-Australis telescope facility at Mount Kent Observatory in Queensland, Australia \cite{2019PASP..131k5003A}.

\subsubsection*{Astrometry}

TOI-201 has been observed by the astrometric space missions \textit{Hipparcos} and \textit{Gaia}, active between 1989 to 1993 and 2014 to 2025 respectively. This allows us to utilize cross-calibrated proper motion data from \textit{Hipparcos-Gaia} astrometry \cite{Kervella2019, Brandt2018} to constrain the reflex motion caused by TOI-201~c over a $\sim$25-year baseline.

We extracted the proper motion data for TOI-201 from the \textit{Gaia}~EDR3 version of the \textit{Hipparcos-Gaia} Catalog of Accelerations \cite{Brandt2021}. In this catalog, the default linear proper motion hypothesis has a $\chi^2$ goodness-of-fit statistic of 40, which is one of the highest values for any confirmed TESS planetary system. In physical units, this is equivalent to a net change in tangential velocity of $134\pm15$~m~s$^{-1}$ between the \textit{Gaia} proper motion and the mean proper motion in the interval between \textit{Hipparcos} and \textit{Gaia} observations. As shown in Fig.~\ref{fig:ttvs_astrometry}, the observed astrometric acceleration is consistent with the signal expected from TOI-201~c.

\subsubsection*{Stellar characterization}

We use the effective temperature, surface gravity, and metallicity from the TESS Input Catalog along with the Gaia DR3 parallax and magnitudes (G, BP, and RP), Two Micron All Sky Survey (2MASS) magnitudes (J, H, and KS), and Wide-field Infrared Survey Explorer (WISE) magnitudes (W1, W2, and W3) to perform an isochrone fit to constrain further the spectroscopic parameters and derive the physical parameters of the host star. The spectroscopic parameters, parallax, and magnitudes are used as priors to determine the goodness of fit. We use the \texttt{isochrone} package \cite{morton15} to generate the isochrone models used to sample the stellar parameters and find the best-fit parameters by using a Markov Chain Monte Carlo (MCMC) routine using the \texttt{emcee} package \cite{dfm13}. The routine consists of 40 independent walkers each taking $5\times10^4$ steps, of which the first 2000 are discarded as burn-in. We find that the host star is a relatively young F star, with an age of $666^{+673}_{-442}$ Myr. The fitted spectroscopic parameters and derived physical parameters, including stellar age, of the host star are reported in Table \ref{tab:star_parameters}.

We use a Generalized Lomb Scargle periodogram \cite{zechmeister2009} to search for periodic stellar variability signals in the TESS light curve after masking out all transits. We detect no consistent period in the TESS photometry, with statistically significant periodicities ranging from less than 1 to more than 10 days depending on the sector analyzed.

\subsubsection*{Statistical validation of the super-Earth}

We rule out false-positive scenarios and statistically validate the super-Earth using the \texttt{triceratops} package \cite{triceratops1, triceratops2}, including the contrast curve from archival SOAR high-resolution imaging to provide additional constraints on the stellar companions generated. We calculate a false-positive probability and nearby false-positive probability of 0.008 and $1\times10^{-12}$, respectively. Given these values, TOI-201~d is a statistically validated planet.

\subsubsection*{Preliminary RV model}

We initially modeled the radial velocities using the \texttt{radvel} package \cite{2018PASP..130d4504F}. We obtained a preliminary orbital solution for the parameters of the outer companion and a tentative mass measurement for the inner super-Earth. We obtained a somewhat well-constrained period, eccentricity, and mass for the brown dwarf that were consistent with both the modeling incorporating the transits and astrometry and the final full photodynamical modeling. We also obtained a mass of $11\pm4$ $M_{\oplus}$ for the super-Earth, which yielded a physically improbable density of 22 g cm$^{-3}$, or four times Earth's density.

\subsubsection*{Joint model incorporating astrometry}

With the overall system architecture of the TOI-201 system having been determined from RV and transit data, we next perform a joint model incorporating the \textit{Hipparcos-Gaia} astrometry. This model is based on the one developed in \cite{Venner2021} to jointly model RVs and \textit{Hipparcos-Gaia} astrometry, implementing modifications for handling multiplanet systems from \cite{Venner2024}. To our knowledge, TOI-201~c is the first substellar companion to be detected simultaneously in RV, transit, and astrometry data. To incorporate the transit data in this model, we use the \texttt{batman} package \cite{batman} to generate transit models.

Since this model assumes Keplerian dynamics, we cannot straightforwardly account for TTVs arising from inter-planet interactions. These effects have no impact on the astrometry at the level of precision; so for the sake of simplicity, we restrict the included photometric data to a single transit each for TOI-201~b and TOI-201~c. We also choose to omit TOI-201~d from this model since its contribution to RV variability is small and to the astrometry negligible. We assume that TOI-201~b does not significantly contribute to the astrometry, which is reasonable as its orbital period is significantly shorter than the 3-year observing baselines of both \textit{Hipparcos} and \textit{Gaia}~DR3.

This model includes a total of 32 variable parameters, of which seven describe the star and system (stellar mass $M_*$, stellar density $\rho_*$, quadratic limb-darkening coefficients $u_1$, $u_2$, parallax $\varpi$, barycentric proper motions $\mu_{\alpha,\text{bary}}$ and $\mu_{\delta,\text{bary}}$), 10 describe the zero-point offsets and jitter terms for the five RV datasets, and the remaining 15 describe the properties of TOI-201~b and TOI-201~c. These parameters are the orbital period $P$, the mass $M_\text{p}$, the eccentricity and argument of periastron parameterized as $\sqrt{e}\sin\omega$, $\sqrt{e}\cos\omega$, transit time $T_0$, impact parameter $b$, and radius ratio $R_\text{p}/R_*$. For TOI-201~c we have additionally the longitude of node $\Omega$, which is used exclusively for fitting to the astrometry.

We reproduce the posterior parameters from this model in Table~\ref{tab:joint_astrometry} and in Fig.~\ref{fig:joint_astrometry} the corresponding fit to the RVs, \textit{Hipparcos-Gaia} astrometry, and the transit of TOI-201~c. It can be seen in the \textit{Hipparcos-Gaia} astrometry that the high significance of proper motion the nonlinearity reported in the \textit{Hipparcos-Gaia} Catalog of Accelerations \cite{Brandt2021} arises in large part from the coincidence of the \textit{Gaia} observations with the previous periastron passage of TOI-201~c, which occurred at BJD~$2457241\pm47$ ($\sim$August 2015; compare \cite{Venner2022}). As a result, the astrometry helps to provide a robust constraint on the orbital period of the outer companion. For TOI-201~c, we find key parameters of $P=2834^{+43}_{-38}$~d, $e=0.610^{+0.047}_{-0.041}$, $\omega=98.0\pm6.2$~deg, and $M_\text{p}=15.4^{+1.0}_{-0.8}~M_J$. The bulk of the posterior constraint on the orbital inclination comes from the transit, rather than the astrometry ($i=89.916^{+0.044}_{-0.022}$~deg); this means that the main degree of freedom constrained by the astrometry is the longitude of node, which we uniquely determine to be $\Omega=212\pm11$~deg.

\begin{figure}
\centering
\includegraphics[width=\textwidth]{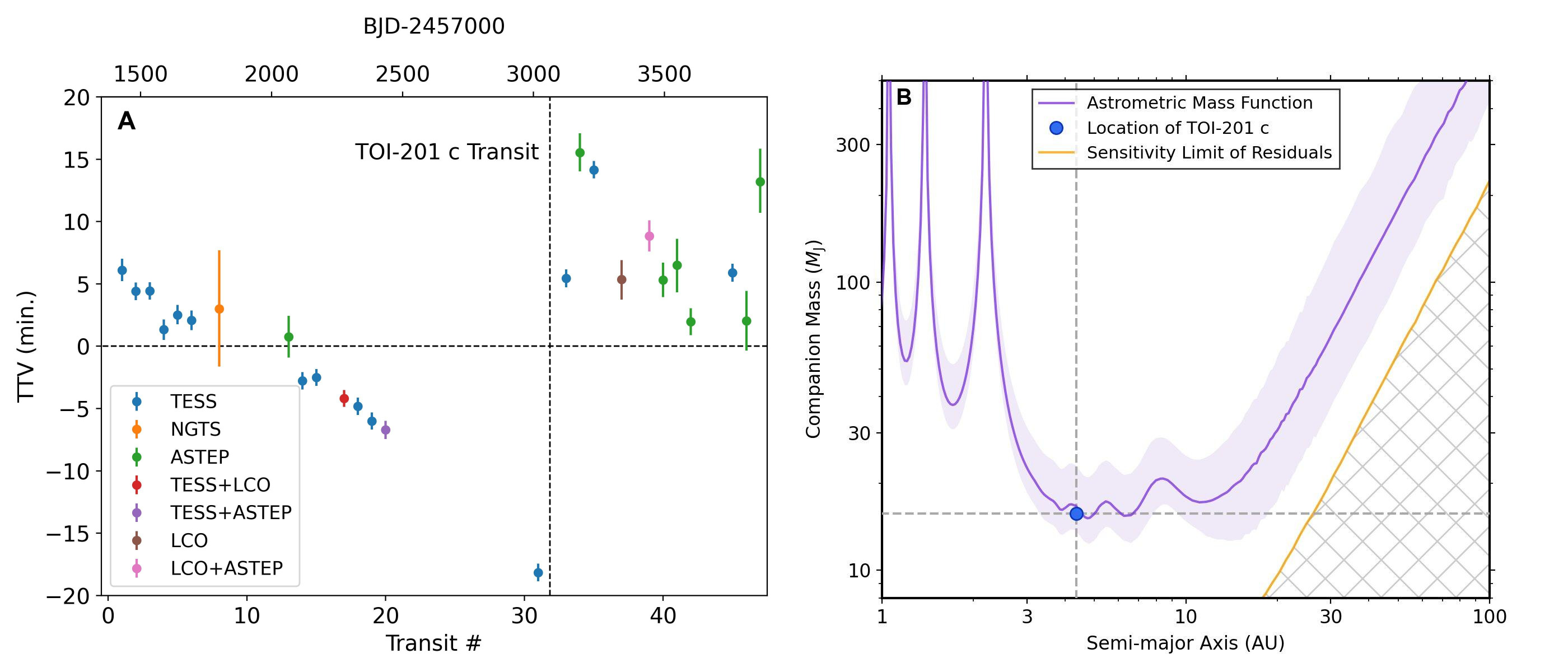}
\caption{\textbf{Evidence for TOI-201~c from TTVs and astrometry.} (\textbf{A}) TTVs for TOI-201~b from TESS and ground-based facilities showing a gradual decline followed by a sudden discontinuity at the time of the outer companion's transit. (\textbf{B}) The astrometric acceleration observed in \textit{Hipparcos-Gaia} astrometry is consistent with the properties of the $\approx$15~$M_J$ outer companion, and otherwise places limits on more massive companions in the system.}
\label{fig:ttvs_astrometry}
\end{figure}

Beyond the detection of TOI-201~c, the $\sim$25 year long temporal baseline of the \textit{Hipparcos-Gaia} astrometry allows us to place limits on the presence of other massive companions in the system. Subtracting our best-fit proper motion model, the 3-sigma upper limit on the remaining \textit{Gaia} tangential velocity anomaly is $<$45~m~s$^{-1}$. We show the mass detection limits from this constraint in Fig.~\ref{fig:ttvs_astrometry}, where companions above the orange line are notionally excluded. In reality, the residual of the astrometric model for TOI-201~c is liable to over-constraint due to the limited scope of the astrometric data, so this detection limit is likely to be optimistic; nonetheless, we believe it is reasonable to infer from the astrometric constraints that stellar-mass companions ($\gtrsim$80~$M_J$) to TOI-201 can be largely ruled out for projected separations within $\lesssim$50~AU.

\subsubsection*{Photodynamical analysis of photometry and RVs}

Since TOI-201~b shows strong TTVs induced by the periastron passage of an eccentric outer companion, similar to Kepler-419~b \cite{2014ApJ...791...89D,2018A&A...615A..90A}, Kepler-448~b and Kepler-693~b \cite{2017AJ....154...64M}, we performed a joint photodynamical analysis of the \tess photometry, the ground-based photometry, and the RVs. The analysis was done using \pyttv following the methodology described in \cite{2023A&A...675A.115K} and \cite{2024ApJ...971L..28K}, assuming a three-planet configuration. We modeled the light curves corrected using the CBV corrector (described above) for Sectors 8 and 64, the \tess SAP light curves with the 2-min cadence for Sectors 1 to 7, 10 to 13, and the 20-s cadence for the remaining sectors. The ASTEP ground-based photometry has been binned to 1.5 minutes. The model is parametrized as described in \cite{2023A&A...675A.115K}. The model is parameterized using the sampling parameters $\sqrt{e}\cos{\omega}$ and $\sqrt{e}\sin\omega$ with half-normal priors on the orbital eccentricities. Since the \texttt{radvel} analysis gives an unrealistically high mass for the innermost planet, we carried out two photodynamical analyses assuming different priors on the planet mass. In the first case, we set a uniform prior on the $\log_{10}$ planet mass from -5 to -4.1, and in the second case, we set a normal prior on the $\log_{10}$ planet mass with a mean of -5.07 and standard deviation of 0.15. The latter one was calculated using the \texttt{spright} mass-radius relation from \cite{Parviainen2023b}. We considered two scenarios for the prior set on the longitude of the ascending nodes. First, we carried out an analysis fixing $\Omega_{d}$ to zero and setting uniform priors on $\Omega_{b}$ and $\Omega_{c}$. Second, we carried out an analysis with a normal prior, $\mathcal{N}(212,11)$, on $\Omega_{c}$, as determined from the astrometry in the preceding analysis, and the longitudes of the ascending nodes of both inner planets were set free. In addition to the main photodynamical analysis, we tested for mass-eccentricity degeneracies by carrying out several photodynamical analysis scenarios assuming different priors on planet masses and eccentricities following \cite{2017AJ....154....5H}. In particular, we used their default prior (log-uniform in planet masses and uniform in eccentricities) and their high-mass prior (uniform in planet masses and log-uniform in eccentricities), which pull the solution toward opposite ends of the degeneracy \cite{2024A&A...688A.211L}. The model parameters and their priors for the main analysis are listed in Table~\ref{tab:pyttv_fitting}.

The PyTTV photodynamical code simultaneously models photometric and RV data using \rebound \cite{rein2012rebound, rein2015whfast, Rein2019a} for dynamical integration. It incorporates general relativity effects through \reboundx \cite{tamayo2020reboundx} and accounts for the light travel time effect \cite{1952ApJ...116..211I}. Transit modeling is performed using \pytransit \cite{Parviainen2015, Parviainen2020a, Parviainen2020b}.
The analysis begins with global optimization using the differential evolution algorithm \cite{Storn1997b, Price2005}. This optimization is followed by MCMC sampling, starting from the global optimization results using the \emcee sampler \cite{dfm13}. Correlated photometric noise is modeled as a Gaussian process, implemented using the \celerite package \cite{Foreman-Mackey2017}. 

\begin{figure}
\centering
\includegraphics[width=\textwidth]{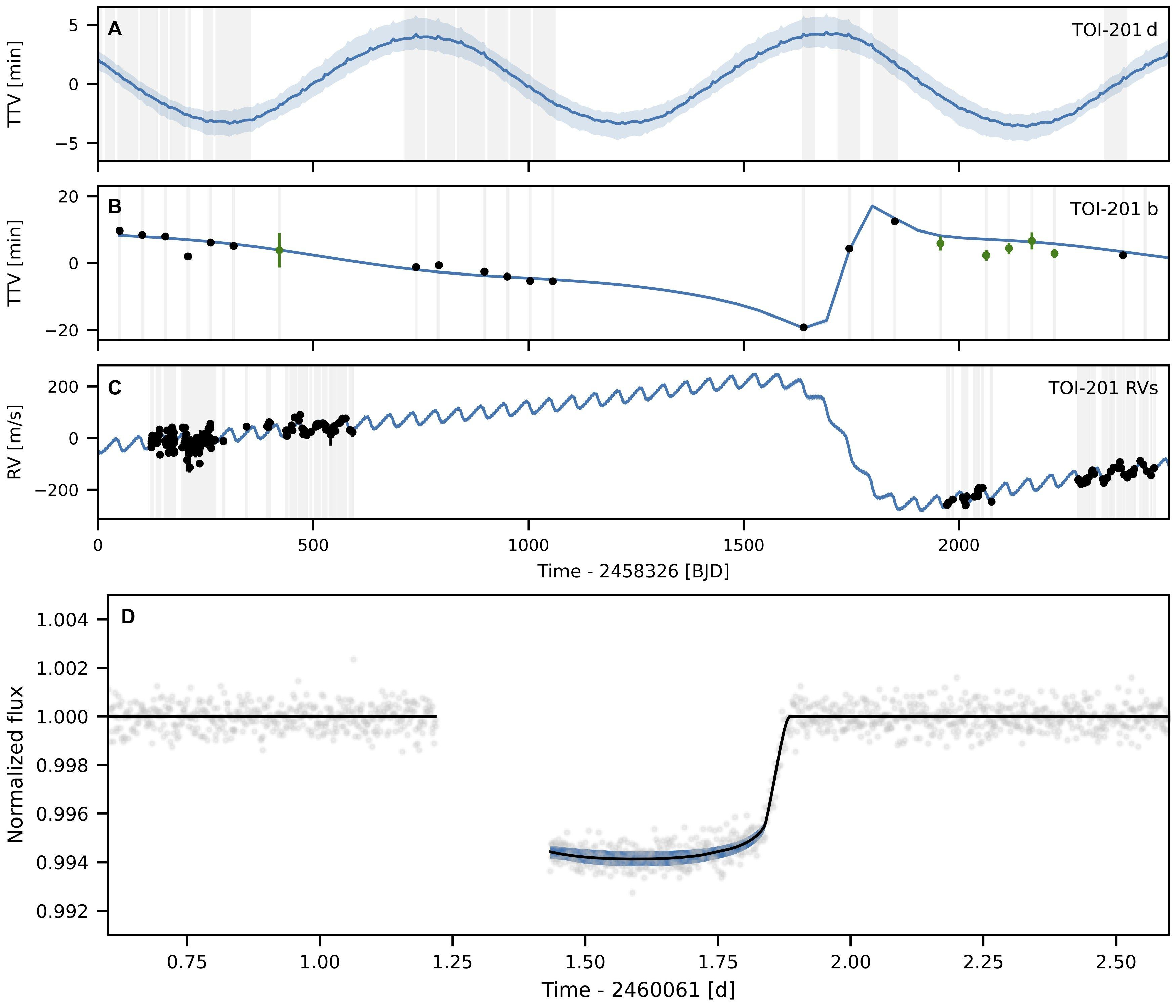}
\caption{\textbf{Photodynamical analysis of TTVs and RVs for TOI-201.} (\textbf{A}) Best-fit TTV model for TOI-201~d. (\textbf{B}) TTVs of TOI-201~b from TESS (black points) and ground-based facilities (green points) and best-fit model. (\textbf{C}) RV data and best-fit RV model. (\textbf{D}) Transit and best-fit model for TOI-201~c. All the panels show the model posterior median as a blue line and the 1-$\sigma$ posterior uncertainties as light blue shading. In (B) and (C), the uncertainties are smaller than the line width.}
\label{fig:data-and-models}
\end{figure}

\begin{figure}
\centering
\includegraphics[width=\textwidth]{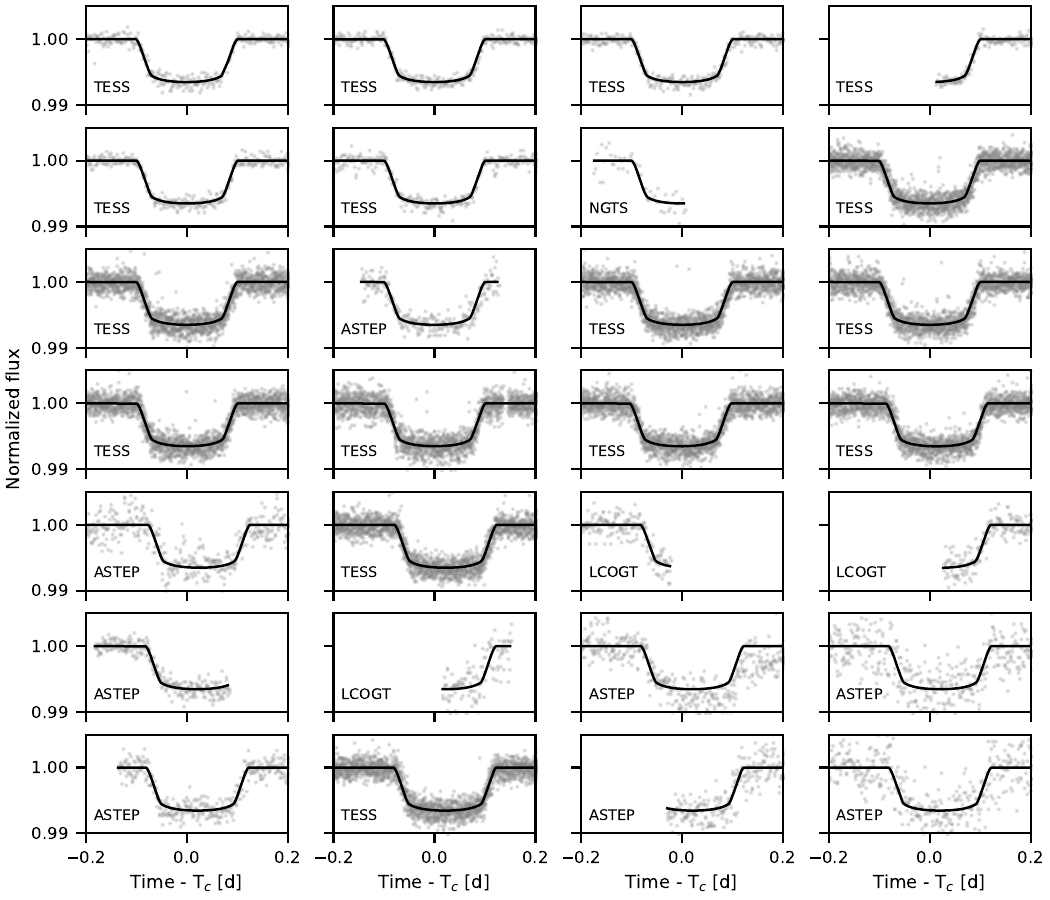}
\caption{\textbf{Data and individual transits of TOI-201~b in the photodynamical model.} The transits are shown in chronological order.}
\label{fig:inditransit_201b}
\end{figure}

We show the modeled TTVs for TOI-201~d and TOI-201~b and the modeled RVs for the system in Fig.~\ref{fig:data-and-models}. In the second panel, we also show the measured TTVs for TOI-201~b, as well as the RV measurements in the lowest panel. The individual transits are shown in Fig.~\ref{fig:inditransit_201b} for TOI-201~b.
The phase-folded plot for TOI-201~d is shown in Fig.~\ref{fig:inditransit_201c}. The photodynamical analysis with the uniform prior on the innermost planet's mass leads to a mass estimate that agrees with the values from the \texttt{radvel} analysis. Since this mass value corresponds to an unrealistically high planet density, we report the solution using the normal prior on the planet mass as the final solution in Table~\ref{tab:pyttv_final}.
The two $\Omega$ scenarios yielded identical posteriors for all the parameters except for the three $\Omega$s.
We note that our posterior solution shows a degeneracy in impact parameter for TOI-201~c and its transit duration (see Fig.~\ref{fig:duration}). This also creates a degeneracy in the impact parameter and transit center time. Observation of a full transit of TOI-201~c is required to solve these degeneracies. The next transit opportunity is 2031-03-26 at 07 UT with an uncertainty of 21 hours. The longitudes of the ascending nodes of both TOI-201~b and TOI-201~c are nearly but not exactly aligned, leading to a mutual inclination of $I_\mathrm{bc}=13\pm2$~deg. There is no significant evidence for nonzero mutual inclinations between the innermost planet and the two outer companions. The limb-darkening parameter, $q_2$, is not well constrained, and therefore we report only the 95th percentile of its posterior distribution. 

\newpage


\begin{figure}
\centering
\includegraphics[width=\linewidth]{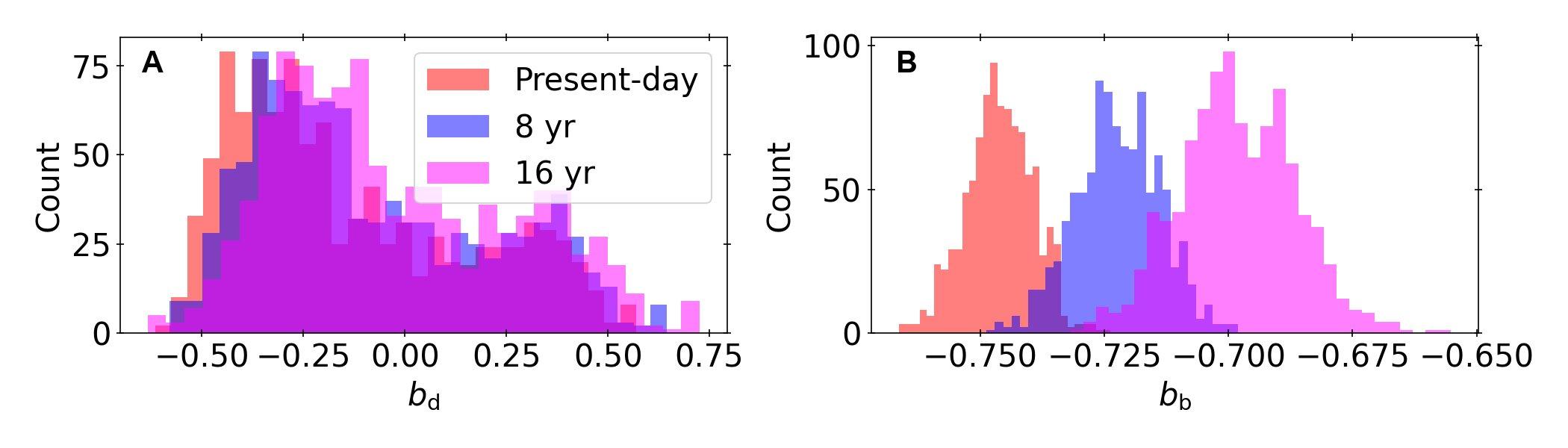}
\caption{\textbf{Short-term evolution of the impact parameters of the two inner planets for 1000 integrations of the posteriors.} (\textbf{A}) The impact parameter of TOI-201~d at the present-day and 8 and 16 years later. (\textbf{B}) The same as (A) but for TOI-201~b.}
\label{fig:impact-param-hists}
\end{figure}

\newpage



\begin{table} 
	\centering
	\caption{\textbf{TOI-201 stellar and planetary properties.} The stellar parameters are derived from our isochrone fit, while the planet parameters are derived from our joint photodynamical-RV fit. The longitude of the ascending node for TOI-201 c comes from our transit-RV-astrometry fit, while the values for b and d come from combining the value for c and the relative longitudes from the photodynamical modeling.}
	\label{tab:main_params} 
	\renewcommand{\arraystretch}{0.575} 
	\begin{tabular}{lcc} 
		\\
		\hline
		\textbf{Stellar Parameters} & \textbf{Value} & \textbf{Reference} \\
        Mass (M$_\odot$) & $1.32^{+0.02}_{-0.04}$ & This work \\
        Radius (R$_\odot$) & $1.31 \pm 0.01$ & This work \\
        Luminosity (L$_\odot$) & $2.61 \pm 0.12$ & This work \\
        Effective temperature (K) & $6423^{-90}_{+86}$ & This work \\
        Age (Gyr) & $0.666^{+0.673}_{-0.442}$ & This work \\
		\hline
		  \textbf{TOI-201~b Parameters} & & \\
		Orbital period (days) & $52.9786\pm0.0001$ & This work \\
		Time of inferior conjunction (BJD) & $2458376.0521\pm0.0002$ & This work \\
        Semi-major axis (AU) & $0.303\pm0.002$ & This work \\
        Inclination ($^\circ$) & $91.18\pm0.02$ & This work \\
        Eccentricity & $0.275\pm0.009$ & This work \\
        Argument of periastron (deg) & $83\pm2$ & This work \\
        Longitude of ascending node (deg) & $198\pm10$ & This work \\
        Radius (Earth radii) & $11.4\pm0.1$ & This work \\
        Mass (Earth masses) & $164\pm5$ & This work \\
        Density (g cm$^{-3}$) & $0.61\pm0.02$ & This work \\
		\hline
        \textbf{TOI-201~c Parameters} & & \\
        Orbital period (days) & $2890\pm20$ & This work \\
		Time of inferior conjunction (BJD) & $2460062.59\pm0.02$ & This work \\
        Semi-major axis (AU) & $4.37\pm0.04$ & This work \\
        Inclination ($^\circ$) & $89.92^{+0.10}_{-0.04}$ & This work \\
        Eccentricity & $0.651\pm0.006$ & This work \\
        Argument of periastron (deg) & $96\pm2$ & This work \\
        Longitude of ascending node (deg) & $211\pm11$ & This work \\
        Radius (Earth radii) & $10.4\pm0.3$ & This work \\
        Mass (Earth masses) & $4990\pm100$ & This work \\
        Density (g cm$^{-3}$) & $24\pm2$ & This work \\
		\hline
        \textbf{TOI-201~d Parameters} & & \\
        Orbital period (days) & 5.84889 $\pm$ 0.00009  & This work \\
		Time of inferior conjunction (BJD) & $2458374.032\pm0.003$ & This work \\
        Semi-major axis (AU) & $0.0698\pm0.0005$ & This work \\
        Inclination ($^\circ$) & $91.7\pm1.6$ & This work \\
        Eccentricity & $0.3\pm0.1$ & This work \\
        Argument of periastron (deg) & $340\pm80$ & This work \\
        Longitude of ascending node (deg) & $175\pm20$ & This work \\
        Radius (Earth radii) & $1.39\pm0.07$ & This work \\
        Mass (Earth masses) & $5.8\pm2$ & This work \\
        Density (g cm$^{-3}$) & $11\pm4$ & This work \\
		\hline
        Inclination between b and d (deg) & $28.0^{+10.6}_{-14.2}$ & This work \\
        Inclination between b and c (deg) & $13.4^{+2.0}_{-2.3}$ & This work \\
        Inclination between c and d (deg) & $41.8^{+10.9}_{-16.4}$ & This work \\
        \hline
	\end{tabular}   
\end{table}

\begin{table}[h!]
\renewcommand{\arraystretch}{0.8}
\centering
\caption{\textbf{Ground-based light curve observations of TOI-201.} The transits of TOI-201~c were not detected as the observations were made before the true orbital period was known and these tested shorter orbital periods. The transits of TOI-201~d were not detected as they are too shallow for the instruments used, although we could rule out nearby eclipsing binaries (NEBs) as the source of the events.}
\label{table:ground_phot}
\begin{tabular}{lllll}
\hline
\hline
Telescope        & Planet & Date         & Filter(s) & Comments                                 \\
\hline
NGTS             & b      & 2019-09-19   & NGTS$^{1}$    & Ingress                                   \\
LCOGT-CTIO       & b      & 2021-01-09   & Y$^{2}$       & Full transit                              \\
LCOGT-SSO        & b      & 2023-12-04   & Y       & Ingress                                   \\
LCOGT-SAAO       & b      & 2023-12-04   & Y       & Egress                                    \\
LCOGT-SAAO       & b      & 2024-03-19   & Y       & Egress                                    \\
LCOGT-SSO        & b      & 2025-02-01   & Y       & Out-of-transit \\
ASTEP            & b      & 2020-06-11   & None      & Egress                                    \\
ASTEP            & b      & 2021-06-17   & None      & Egress                                    \\
ASTEP            & b      & 2023-06-28   & $\sim$ BP$^{3}$, $\sim$ RP$^{4}$   & Full transit                                  \\
ASTEP            & b      & 2024-03-19   & $\sim$ BP, $\sim$ RP   & Ingress                                   \\
ASTEP            & b      & 2024-05-11   & $\sim$ BP, $\sim$ RP   & Full transit                              \\
ASTEP            & b      & 2024-07-03   & $\sim$ BP, $\sim$ RP   & Full transit                              \\
ASTEP            & b      & 2024-08-25   & $\sim$ BP, $\sim$ RP   & Full transit                              \\
ASTEP            & b      & 2025-03-25   & $\sim$ BP, $\sim$ RP   & Egress                                    \\
ASTEP            & b      & 2025-05-17   & $\sim$ BP, $\sim$ RP   & Full transit                                  \\
\hline
LCOGT, PEST, HATPI, & c & 2023-11-29 to & Various & Various periods ruled out \\
Unistellar Network, & & 2024-12-20 & & (see text) \\
Hazelwood Observatory & & & & \\
\hline
LCOGT-CTIO       & d      & 2020-11-04   & zs$^{5}$      & No NEBs          \\
LCOGT-SAAO       & d      & 2021-01-01   & zs      & No NEBs          \\

\hline
\hline
\multicolumn{5}{l}{$^1$Custom filter with bandpass 5200--8900\,\AA \:$^2$Pan-STARRS Y-band ($\lambda_{\rm c} = 10040$\,\AA, ${\rm Width} =1120$\,\AA)} \\
\multicolumn{5}{l}{$^3$Similar to $Gaia-\rm{BP}$ band \:$^4$Similar to $Gaia-\rm{RP}$ band} \\
\multicolumn{5}{l}{$^5$ Pan-STARRS $z_s$ band ($\lambda_{\rm c} = 8700$\,\AA, ${\rm Width} =1040$\,\AA)} \\
\end{tabular}

\end{table}


\clearpage 

%
\bibliography{toi201_bib, browndwarf_library} 
\bibliographystyle{sciencemag}

%
%
%
%
%
%


\section*{Acknowledgments}
We acknowledge the use of public TESS data from pipelines at the TESS Science Office and at the TESS Science Processing Operations Center. Resources supporting this work were provided by the NASA High-End Computing (HEC) Program through the NASA Advanced Supercomputing (NAS) Division at Ames Research Center for the production of the SPOC data products.
This work makes use of observations from the LCOGT network. Part of the LCOGT telescope time was granted by NOIRLab through the Mid-Scale Innovations Program (MSIP). MSIP is funded by NSF. This research has made use of the Exoplanet Follow-up Observation Program (ExoFOP; DOI: 10.26134/ExoFOP5) website, which is operated by the California Institute of Technology, under contract with the National Aeronautics and Space Administration under the Exoplanet Exploration Program. We gratefully acknowledge access to computational resources through the MIT Engaging cluster at the Massachusetts Green High Performance Computing Center (MGHPCC) facility and the MIT SuperCloud and Lincoln Laboratory Supercomputing Center \cite{reuther2018interactive}.
The ASTEP team thanks the dedication and technical support of the entire French Polar Agency (IPEV). We also wish to thank the technical staff at Concordia Station, and give a particular recognition to the work and efforts produced by the entire wintering crew at Concordia to ensure a continuity of operations throughout each Antarctic winter. ASTEP benefited from the support of the French and Italian polar agencies IPEV and PNRA in the framework of the Concordia station program and from OCA, INSU, Idex UCAJEDI (ANR-15-IDEX-01) and ESA through the Science Faculty of the European Space Research and Technology Centre (ESTEC).

\paragraph*{Funding:}
Funding for the TESS mission is provided by NASA’s Science Mission Directorate. We acknowledge the use of public TESS data from pipelines at the TESS Science Office and at the TESS Science Processing Operations Center. Resources supporting this work were provided by the NASA High-End Computing (HEC) Program through the NASA Advanced Supercomputing (NAS) Division at Ames Research Center for the production of the SPOC data products.
This paper was supported by the National Science Foundation under Grant No. 2306391. 
D.D. acknowledges support from the TESS Guest Investigator Program grant 80NSSC23K0769. D.D. thanks the Kavli Institute for Theoretical Physics (funded by grant NSF PHY-2309135) for hospitality while parts of this work were completed. 
The authors acknowledge support from the Swiss NCCR PlanetS and the Swiss National Science Foundation. This work has been carried out within the framework of the NCCR PlanetS supported by the Swiss National Science Foundation under grants 51NF40182901 and 51NF40205606. J.K. acknowledges support from the Swedish Research Council (Project Grant 2017-04945 and 2022-04043) and of the Swiss National Science Foundation under grant number TMSGI2\_211697.
K.A.C. acknowledges support from the TESS mission via subaward s3449 from MIT.
This research received funding from the European Research Council (ERC) under the European Union's Horizon 2020 research and innovation programme (grant agreement n$^\circ$ 803193/BEBOP), and from the Science and Technology Facilities Council (STFC; grant n$^\circ$ ST/S00193X/1, ST/W002582/1, and ST/Y001710/1).
D.Ra. was supported by NASA under award number 80NSSC25M7110.
H.P. acknowledges support from the Spanish Ministry of Science and Innovation with the Ramon y Cajal fellowship number RYC2021-031798-I, and funding from the University of La Laguna and the Spanish Ministry of Universities.
This paper is in part based on data collected under the NGTS project at the ESO La Silla Paranal Observatory. The NGTS facility is operated by the consortium institutes with support from the UK Science and Technology Facilities Council (STFC) projects ST/M001962/1 and ST/S002642/1. We acknowledge financial support from the Agencia Estatal de Investigaci\'on of the Ministerio de Ciencia e Innovaci\'on MCIN/AEI/10.13039/501100011033 and the ERDF “A way of making Europe” through project PID2021-125627OB-C32, and from the Centre of Excellence “Severo Ochoa” award to the Instituto de Astrofisica de Canarias.
P.F. acknowledges financial support from the Severo Ochoa grant CEX2021-001131-S funded by MCIN/AEI/10.13039/501100011033. P.F. is also funded by the European Union (ERC, THIRSTEE, 101164189). Views and opinions expressed are however those of the author(s) only and do not necessarily reflect those of the European Union or the European Research Council. Neither the European Union nor the granting authority can be held responsible for them.
M.L. acknowledges support of the Swiss National Science Foundation under grant number PCEFP2\_194576
Funding for K.B. was provided by the European Union (ERC AdG SUBSTELLAR, GA 101054354).
L.A.S. was supported during this work by the NASA Citizen Science Seed Funding Program via grant No. 80NSSC22K1130 and the NASA Exoplanets Research Program via grant 80NSSC24K0165, which also support the UNITE (Unistellar Network Investigating TESS Exoplanets) program.

\paragraph*{Author contributions:}
I.M. contributed to the planet detection, analysis, and led the writing for the paper.
S.U.M. coordinated CORALIE and HARPS observations, contributed to the data reduction for CORALIE and HARPS, and contributed to the analysis of the RVs.
D.L. led the development and analysis of the dynamical simulations and contributed to the writing of the manuscript.
D.D. contributed to the interpretation of the results and writing of the manuscript.
J.K. led the analysis for the photodynamical modeling and contributed to the writing of the manuscript.
A.V. led the analysis incorporating the \textit{Hipparcos-Gaia} astrometry and the relevant sections of text.
K.A.C is the lead for TFOP SG1, coordinated LCO observations, and contributed to the LCO data reduction.
A.H.T. contributed to the astrometric identification.
A.H.T. and T.G. led the observing of ASTEP.
A.P. contributed to the analysis and interpretation of the TTVs.
T.C. contributed to the interpretation of the brown dwarf as it relates to the known sample of transiting brown dwarfs. 
S.M. and T.H. contributed to the development and analysis of the dynamical simulations and contributed to the writing of the manuscript.
H.P. contributed to the photodynamical modeling.
H.P.O. led the MonoTools analysis used to schedule ground-based observations.
D.Ra. led the reprocessing of the TESS data for Sectors 8 and 64.
T.A.B contributed to the astrometric identification.
S.B., F.B., R.D., P.F., M.L., S.U., and P.W. contributed to the acquisition, data reduction, and interpretation of the CORALIE and HARPS data.
L.A., A.A., M.B., P.B., V.D., D.M., F.-X.S., and O.S. contributed to the obtaining and data reduction for ASTEP observations.
K.B., R.P.S., and F.P.W. contributed to the LCO data reduction and photometric extraction.
K.H. contributed time for LCO observations from the Scottish Universities Physics Alliance (SUPA).
F.M and E.P. contributed LCO observing time and participated on the manuscript writing.
R.S. contributed time for LCO observations from the South African Astronomical Observatory (SAAO).
A.S. is the PI for the LCO Key Project.
G.S. contributed to the organization of LCO and data reduction for LCO observations.
C.S. contributed to the LCO and Hazelwood data reduction and photometric extraction. 
J.D.H. led the observing and analysis of HATPI data.
L.A.S. led the observing and analysis of Unistellar Network data.
T.T. contributed to the PEST observations and data reduction.
J.M.J. contributed to the reprocessing of TESS data.
A.B. is a member of the HATPI team.
D.H., D.Ri., and I.T. contributed Unistellar photometric observations. 

\paragraph*{Competing interests:}
D.Ra. is also affiliated with the Center for Astrophysics and Space Astronomy, Department of Astrophysical and Planetary Sciences, University of Colorado Boulder, CO 80309, USA. The authors declare that they have no other competing interests.

\paragraph*{Data, code, and materials availability:}

The TESS photometric observations are available at the Mikulski Archive for Space Telescopes (MAST) at \url{https://exo.mast.stsci.edu}. Ground-based photometric observations from LCOGT, ASTEP, NGTS, PEST, HATPI, Unistellar Network, and Hazelwood Observatory are available at \url{https://exofop.ipac.caltech.edu/tess} under the name TOI-201. CORALIE, HARPS, and FEROS spectra are available at the ESO Science Archive Facility at \url{http://archive.eso.org/cms.html}. All RVs are available as machine-readable files in Data S1-S5. The \textit{Hipparcos-Gaia} astrometry used in this work is drawn from \cite{Brandt2021}. The publicly available codes used in this work are: \texttt{REBOUND}: \url{https://rebound.readthedocs.io/en/latest/}; \texttt{REBOUNDx}: \url{https://reboundx.readthedocs.io/en/latest/}; \texttt{PyTransit}: \url{https://pytransit.readthedocs.io/en/latest/}; \texttt{emcee}: \url{https://emcee.readthedocs.io/en/stable/}; \texttt{radvel}: \url{https://radvel.readthedocs.io/en/latest/}; \texttt{batman}: \url{https://lkreidberg.github.io/batman/docs/html/index.html}; \texttt{Lightkurve}: \url{https://lightkurve.github.io/lightkurve/}; \texttt{MonoTools}: \url{https://github.com/hposborn/MonoTools}; \texttt{triceratops}: \url{https://github.com/stevengiacalone/triceratops}; and \texttt{isochrones}: \url{https://isochrones.readthedocs.io/en/latest/}. No new materials were generated in this work.

\subsection*{Supplementary materials}
Supplementary Text\\
Figs. S1 to S15\\
Tables S1 to S7\\
References \textit{(1-\arabic{enumiv})}\\ 
Data S1-S5


\newpage


\renewcommand{\thefigure}{S\arabic{figure}}
\renewcommand{\thetable}{S\arabic{table}}
\renewcommand{\theequation}{S\arabic{equation}}
\renewcommand{\thepage}{S\arabic{page}}
\setcounter{figure}{0}
\setcounter{table}{0}
\setcounter{equation}{0}
\setcounter{page}{1} 


\begin{center}
\section*{Supplementary Materials for\\ \scititle}

Ismael Mireles$^\ast$ et al.\\
$^\ast$Corresponding author. Email: mirelesi@unm.edu\\
\end{center}

\subsubsection*{This PDF file includes:}
Supplementary Text\\
Figures S1 to S15\\
Tables S1 to S7\\
Captions for Data S1 to S5


\newpage


\subsection*{Supplementary Text}

\subsubsection*{Present-day dynamics}
In this section our goal is to characterize the present-day dynamics of the TOI-201 system. Through a suite of numerical orbit integrations, we show that TOI-201 is a relatively stable system that nonetheless possesses a nonzero chance of chaotic tidal destruction of planet d. We also demonstrate that the TOI-201 system exhibits secular dynamical changes in planetary transits over timescales directly observable by humans. 

To study the present-day dynamics of the TOI-201 system, we sample the system's orbital parameters from the derived posterior distributions summarized in Table \ref{tab:pyttv_final} and integrate using the $N$-body code \texttt{REBOUND} \cite{rein2012rebound}. Specifically, we use the integrator \texttt{WHFast} \cite{wisdom1991symplectic, rein2015whfast} and choose a timestep which is $1/50$th of the innermost planet's orbital period to ensure numerical stability. We conduct two sets of integrations: Sample A consists of 500 simulations, each spanning 1 Myr, while Sample B includes 100 simulations, each for 10 Myr. The sampling of the posterior distributions is independent between Samples A and B.

\begin{figure}
    \centering
    \includegraphics[width=\linewidth]{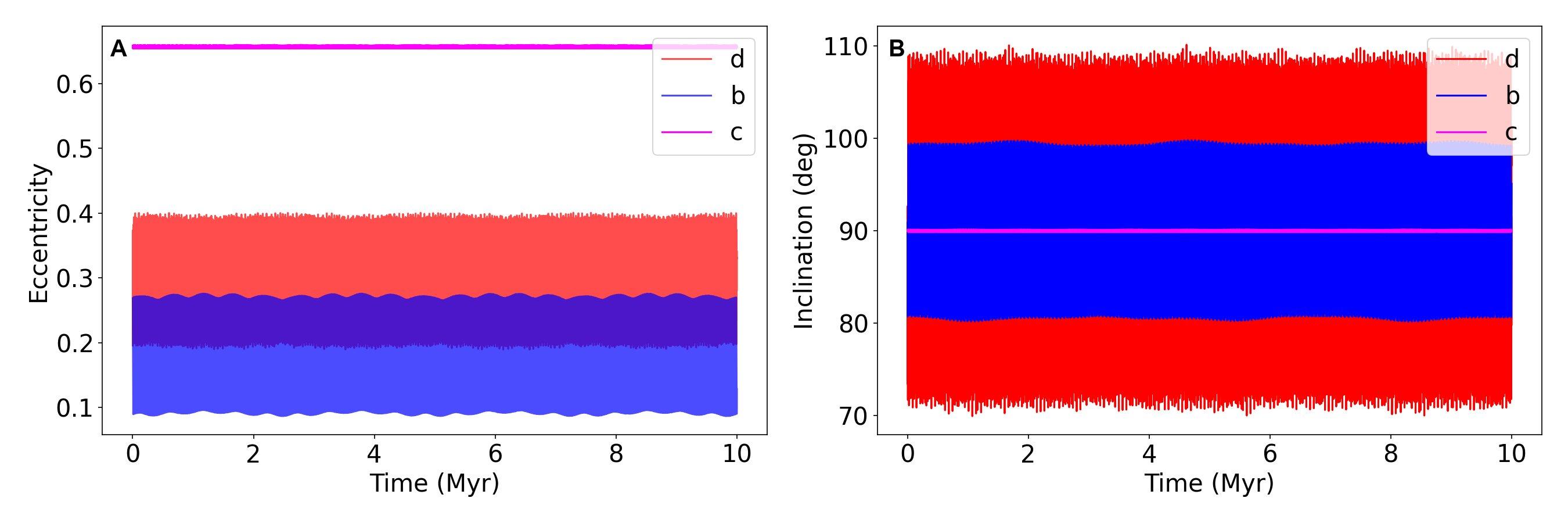}
    \caption{\textbf{Dynamical evolution of the TOI-201 system for a 10 Myr integration.} (\textbf{A}) The evolution of the eccentricities showing minimal changes for TOI-201~c and larger amplitude oscillations for TOI-201~b and d. (\textbf{B}) The evolution of the inclinations showing minimal changes for TOI-201~c, moderate amplitude oscillations for TOI-201~b, and larger amplitude oscillations for TOI-201~d.}
    \label{fig:dynamics-10myr}
\end{figure}

\begin{figure}
    \centering
    \includegraphics[width=\linewidth]{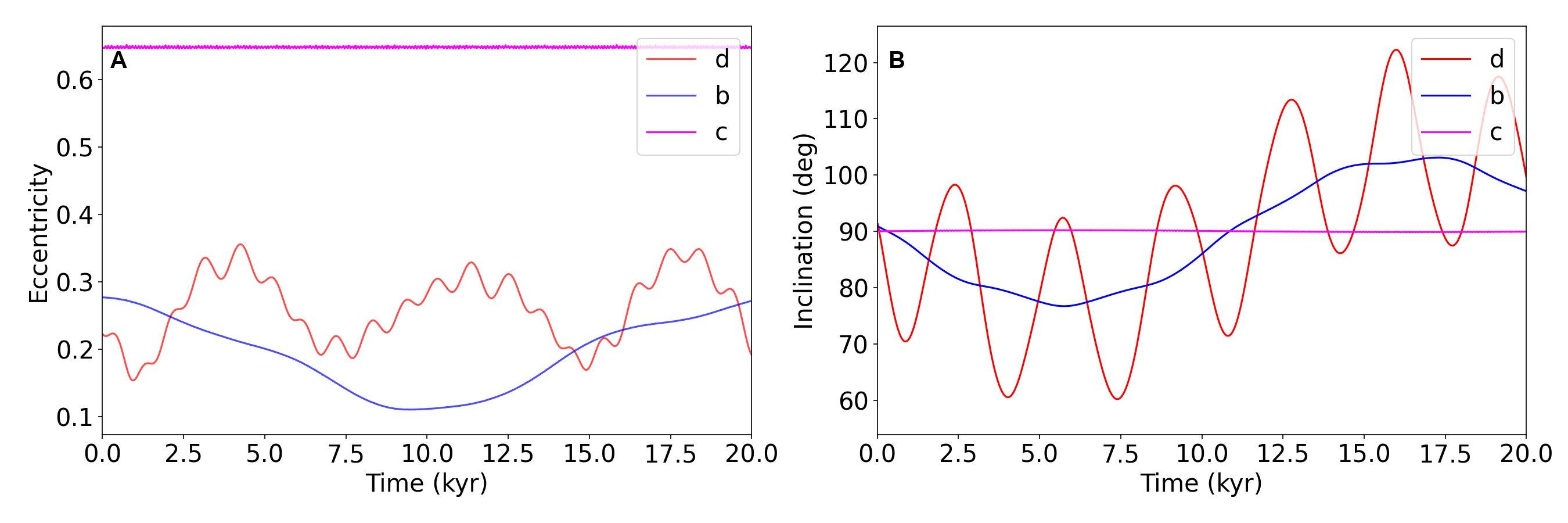}
    \caption{\textbf{Secular interactions within the TOI-201 system over 20 kyr.} (\textbf{A}) The evolution of the eccentricities showing minimal changes for TOI-201~c, mostly long timescale changes for TOI-201~b, and both long and short timescale changes for TOI-201~d. (\textbf{B}) The evolution of the inclinations showing the same patterns as (A).}
    \label{fig:dynamics-20kyr}
\end{figure}

We find in both Samples A and B that the high eccentricities and mutual inclination of TOI-201~b and c result in high-amplitude secular oscillations in the inclination and eccentricity of TOI-201~b and d's orbits (Figures \ref{fig:dynamics-10myr}, \ref{fig:dynamics-20kyr}). These secular interactions result in the excitation of TOI-201~d onto a moderately eccentric ($e \sim 0.2-0.5$) orbit. In Sample A, the mean amplitude of eccentricity oscillations of TOI-201~b and TOI-201~d (defined as $e_{\text{max}} - e_{\text{min}}$) is $0.13$ and $0.31$, respectively. In $\sim 5\%$ of integrations in Sample A, secular interactions result in immediate chaotic evolution of TOI-201~d's eccentricity and subsequent collision with the host star. There are no clear constraints on the posterior distribution if we eliminate the unstable initial conditions. We hypothesize that these systems are in close proximity to secular resonances and that secular analysis may be used to further constrain the posterior distributions \cite{murray1999solar}.

Interestingly, a small proportion of systems ($\sim 3$ \%) from Sample B which are stable on the 100 kyr timescale exhibit chaotic growth of TOI-201~d's eccentricity after $500$ kyr. To determine whether this chaotic behavior may lead to the formation of an ultra-short period planet in a similar scenario to that explored in \cite{petrovich2019ultra}, we carry out additional integrations that account for equilibrium tides and general relativity and result in suppression of high $(e\gtrsim 0.9)$ eccentricities. Specifically, we use the \texttt{gr\_potential} and \texttt{tides\_constant\_time\_lag} \cite{baronett2022stellar} modules in \texttt{REBOUNDx} \cite{tamayo2020reboundx} and choose TOI-201~d to have tidal Love number $k_2 = 0.4$ and quality factor $Q \sim 10^4$, consistent with a sub-Neptune composition \cite{millholland2019obliquity}. We find that these effects are not enough to quench the eccentricity excitations driven by secular chaos, and thus TOI-201~d may (in rare circumstances) undergo tidal disruption on the $\sim$ Myr timescale, unless it is spared by more extreme tidal effects such as chaotic dynamical tides \cite{vick2019chaotic, liveoak2025self}.

Given the strong eccentricity oscillations of TOI-201~b and c, it is reasonable to question whether either or both bodies may be undergoing tidal migration. Assuming $Q \sim 10^5$ and $k_2 = 0.25$ for both TOI-201~b and TOI-201~c, we find that neither body is currently experiencing tidal migration on astrophysically relevant timescales via equation 3 of \cite{levrard2009falling}, assuming upper bounds on both orbital eccentricities of $0.7$ (as motivated by our earlier integration results). We also investigated whether TOI-201~b and c are undergoing tidal migration due to an additional, undetected planet in the system contributing to dynamical excitation. To determine whether the system could host an additional planet between the orbits of TOI-201~b and c, we use a Mean Exponential Growth of Nearby Orbits (MEGNO) map implemented in \texttt{REBOUND} \cite{maffione2011testing}. Specifically, systems whose MEGNO score is $\gtrsim 3$ for a 10 kyr integration are chaotic and inferred to be unstable on astronomical timescales. With this criterion, we find no stable orbits of super-Earth, Saturn, or Jupiter-sized objects between TOI-201~b and c.

\begin{figure}
    \centering
    \includegraphics[width=\linewidth]{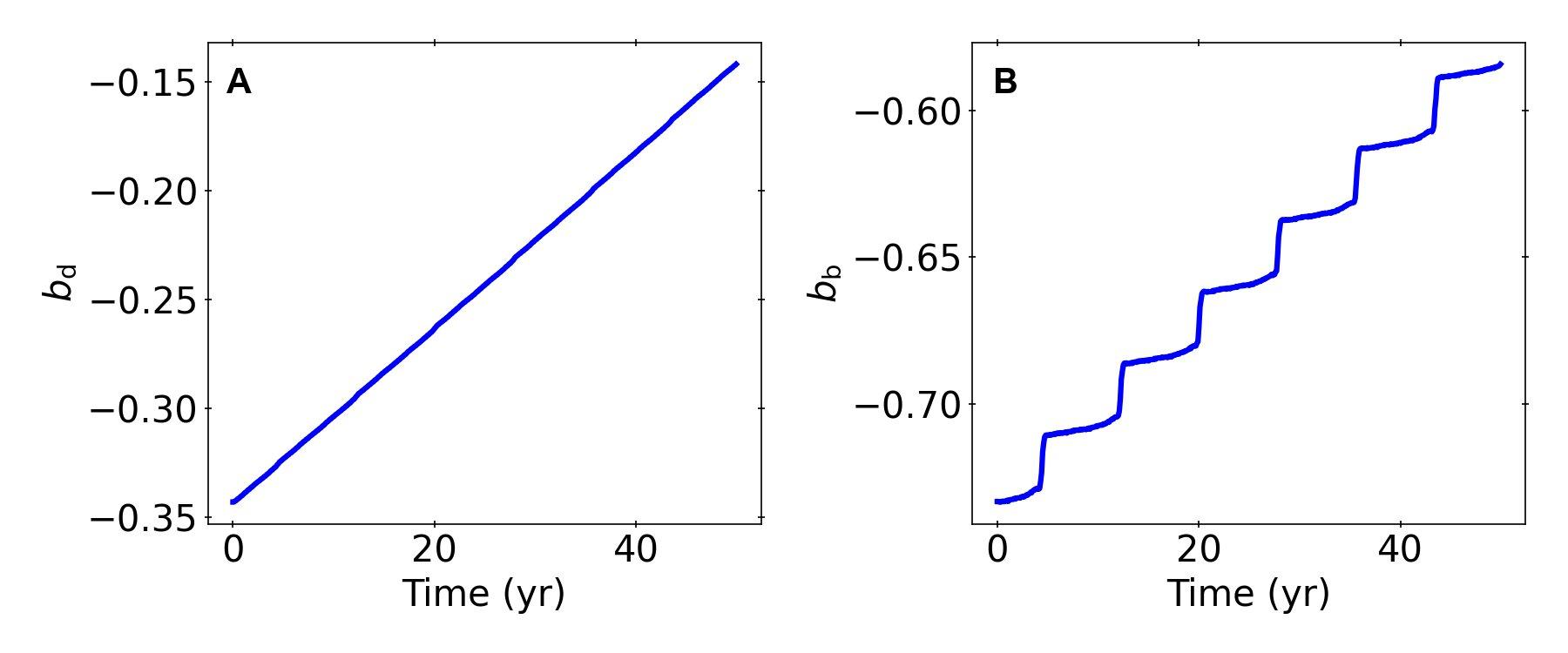}
    \caption{\textbf{Short-term dynamics of the impact parameters of the two inner planets.} (\textbf{A}) The evolution of the impact parameter for TOI-201~d showing a steady linear change. (\textbf{B}) The evolution of the impact parameter for TOI-201~b showing sharp increases after every periastron passage of TOI-201~c.}
    \label{fig:impact-parameter-decade}
\end{figure}

We also find that strong secular and quasi-secular interactions induce meaningful changes in the inclinations of TOI-201~b and d on human-observable ($\sim 10$ yr) timescales. Specifically, we study the short-term evolution of the transit impact parameter $b=a\cos i/R_\star$ for each planet. Each planet is transiting only when its impact parameter satisfies $|b| < 1$. We find that secular interactions between TOI-201~b, c and d result in substantial changes to the impact parameters of TOI-201~b and d, denoted $b_\text{b}$ and $b_\text{d}$, over decade timescales (see Fig. \ref{fig:impact-parameter-decade}). Furthermore, there are periodic impulses in $b_\text{b}$ which correspond to the periastron of TOI-201~c. These quasi-secular perturbations are, in principle, detectable in as soon as 8 years (Fig. \ref{fig:impact-param-hists}).

\begin{figure}
    \centering
    \includegraphics[width=\linewidth]{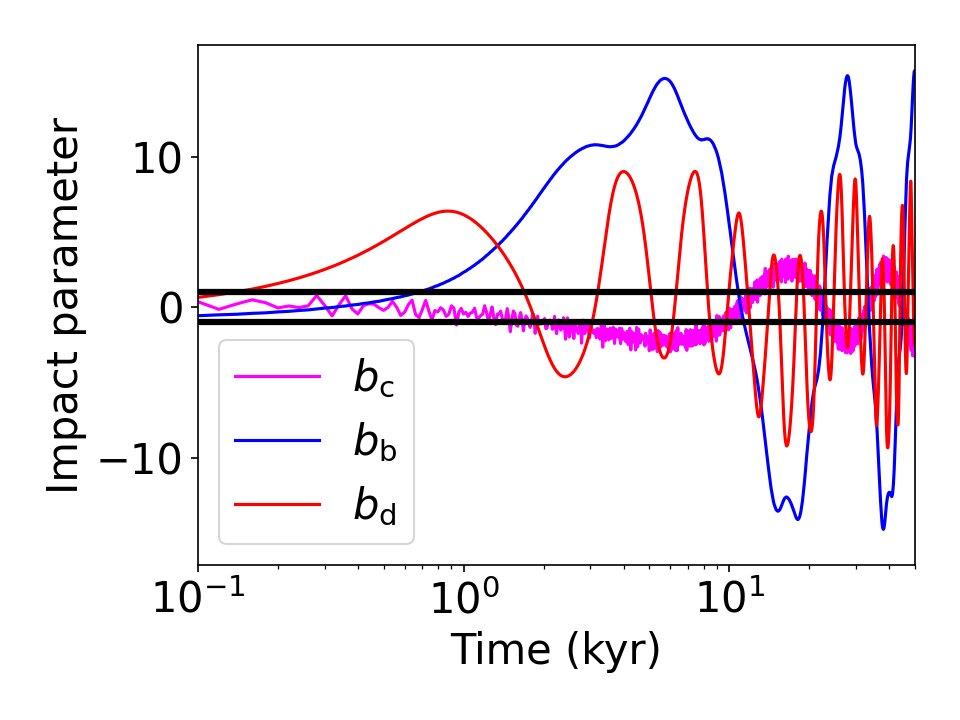}
    \caption{\textbf{Dynamics of the impact parameters of TOI-201~b, c, and d over 50 kyr.} All planets are visible via transit observations only when the impact parameters lie between the black horizontal lines, which indicate $b=\pm 1$.}
    \label{fig:impact-dynamics}
\end{figure}

Among the integrations in Sample A, the mean times for TOI-201~d and TOI-201~b to cease transiting are 275 and 615 yr, respectively. Furthermore, the mean time for the system to return to a cotransiting configuration after one planet's impact parameter exceeds unity is $21$ kyr. The dynamics of each of the planet's impact parameters is shown in Fig.~\ref{fig:impact-dynamics}.

\subsubsection*{Evolution of the TOI-201 system}
Several aspects of the TOI-201 system make its dynamical history difficult to explain given conventional channels of planet formation.  Disk interactions typically only excite eccentricities as high as $e \sim 0.4$ \cite{chen2021wide, baruteau2021observational}, unless the inner edge of the disk is truncated \cite{romanova2023eccentricity}; however, this scenario is in tension with the presence of TOI-201~b and TOI-201~d which are on tighter orbits than TOI-201~c. Here, we evaluate several mechanisms which could explain the architecture of the TOI-201 system, including high-eccentricity migration, stellar fly-bys, planet-planet scattering, and von-Zeipel-Lidov-Kozai (vZLK) cycles.

We find that the existence of TOI-201~d on a tight orbit rules out the possibility of high eccentricity tidal migration (e.g. \cite{mustill2015destruction, petrovich2015hot, socrates2012super}) as a formation channel of TOI-201~b. In particular, we use a MEGNO map in \texttt{REBOUND} and sample regions of parameter space with hypothetical proto-TOI-201~b's on orbits consistent with the planet's observed angular momentum but higher eccentricity. We find that for $e \gtrsim 0.8$, TOI-201~d is ejected on the $\sim 100$ year timescale. A similar analysis, combined with previous established bounds on the tidal capture of brown dwarfs \cite{winter2022forming, bonnell2003there} rules out the possibility of TOI-201~c having undergone high-eccentricity tidal migration or tidal capture in its dynamical history.

\begin{figure}
    \centering
    \includegraphics[width=\linewidth]{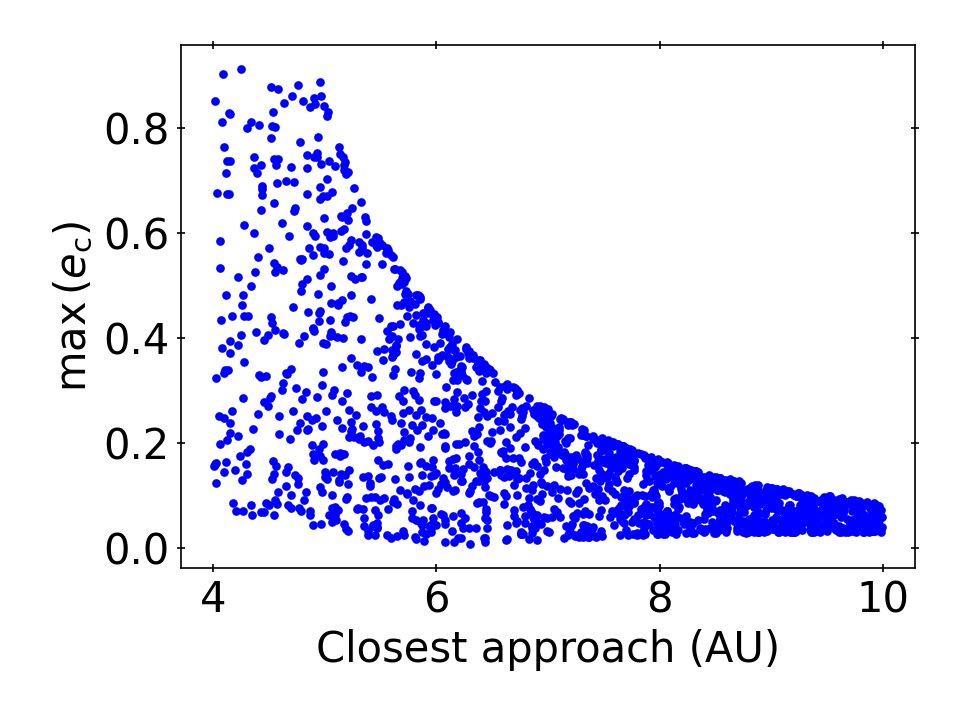}
    \caption{\textbf{Maximum eccentricity of TOI-201~c vs. closest approach of perturbing star in stellar fly-by simulations.} The closest approach must be less than 6 AU in order to reproduce the observed eccentricity.}
    \label{fig:flyby-closest}
\end{figure}

Through a suite of $N$-body simulations, we find that pumping the eccentricities of TOI-201~b and c through a stellar flyby requires a very tight distance of closest approach of ${\sim}4-6$ au (Fig.~\ref{fig:flyby-closest}). To establish whether such flyby distances are feasible in typical cluster environments, we estimate the timescale to achieve such a flyby following $t_{\rm flyby}{\sim}(n_{\star}\sigma_{\star}d^{2}_{\rm flyby})^{-1}$ where $n_{\star}$ is the stellar density, $\sigma_{\star}$ is the velocity dispersion, and $d_{\rm flyby}{\sim}5$ au is the flyby distance. In young open clusters, $n_{\star}{\sim}10$ pc$^{-3}$ and $\sigma_{\star}{\sim}3$ km s$^{-1}$ \cite{bintre08}, which implies that $t_{\rm flyby}{\gg}$ Gyr. We therefore conclude that flybys sufficiently close to excite the eccentricities of TOI-201 b and c are unlikely to occur.

Having shown that flybys are unlikely to excite the eccentricities of TOI-201~b and c, we next consider the possibility of planet-planet scattering as a mechanism for generating the high eccentricities of TOI-201~b and c. Planet-planet scattering is hypothesized to be an important pathway for the sculpting of planetary systems, due to the observed diversity of eccentricities of giant planets and the inferred compactness of configurations formed via core accretion \cite{chatterjee2008dynamical, ford2008origins}.

To evaluate the plausibility of the planet-planet scattering scenario, we carry out a suite of $N$-body simulations. Specifically, we initialize TOI-201~b and introduce a hypothetical TOI-201~e with mass $7\,M_\text{jup}$ and semi-major axis sampled uniformly from $4$ to $4.5$ au. TOI-201~c's orbit is initialized such that it has a mutual Hill spacing $\Delta = 3$ with the hypothetical TOI-201e. We neglect considerations of TOI-201~d to improve simulation runtime. All orbits are initially approximately coplanar and moderately eccentric, with mutual inclinations $\lesssim 3 \deg$ and eccentricities between $0$ and $0.2$. We use the hybrid integrator \texttt{mercurius} \cite{rein2019hybrid} to ensure numerical stability despite close encounters, and carry out 4000 integrations of the initial conditions for $100$ kyr each.

\begin{figure}
    \centering
    \includegraphics[width=\linewidth]{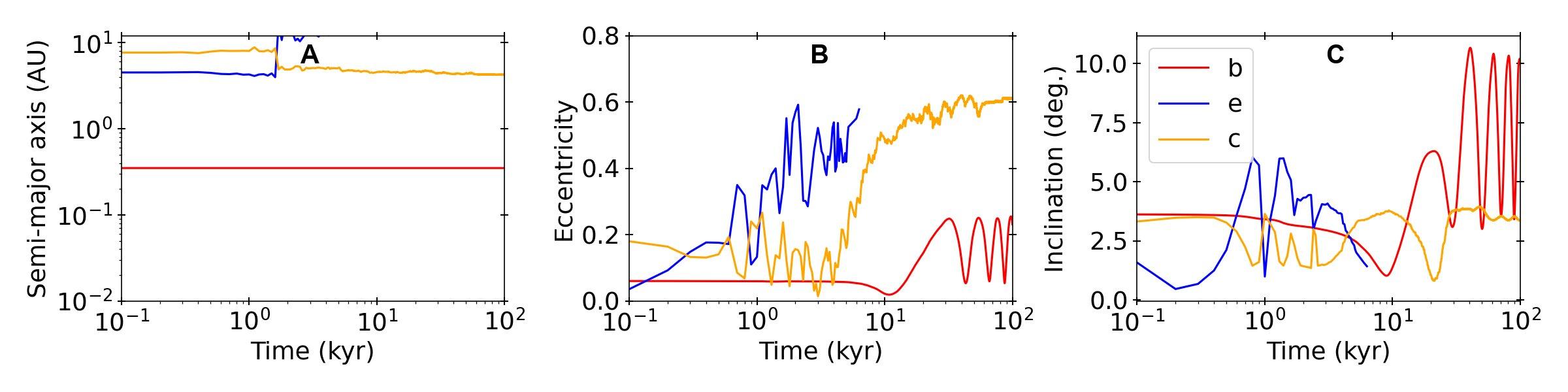}
    \caption{\textbf{Dynamical evolution of planet-planet scattering simulation which reproduces the observed architecture of the TOI-201 system.} (\textbf{A}) Evolution of the semi-major axes showing the injected planet being ejected and TOI-201~c moving slightly inward. (\textbf{B}) Evolution of the eccentricity showing the growth of TOI-201~c's eccentricity and start of oscillations as the injected planet is ejected. (\textbf{C}) Evolution of the inclinations.}
    \label{fig:scattering-example}
\end{figure}

\begin{figure}
    \centering
    \includegraphics[width=\linewidth]{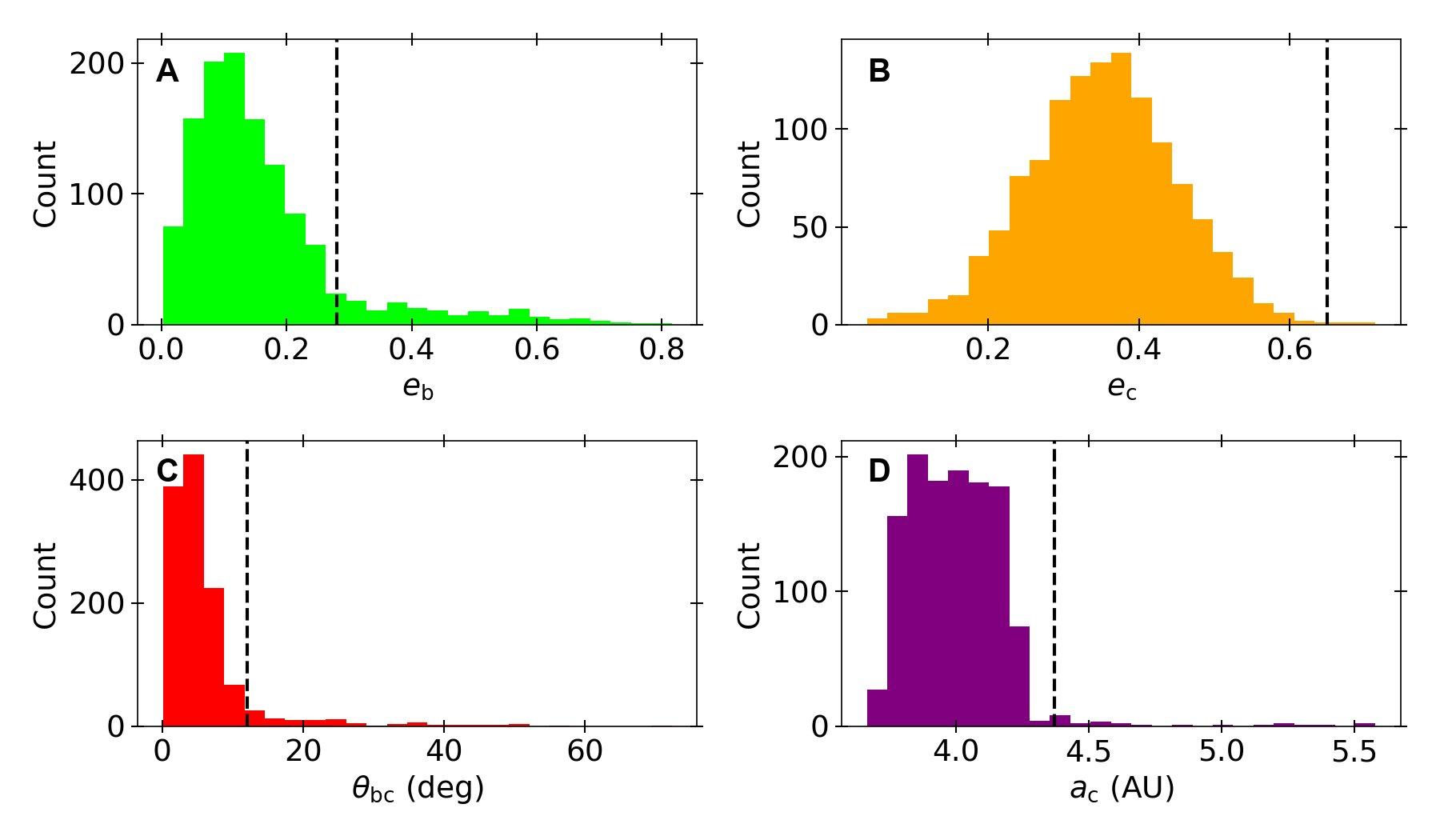}
    \caption{\textbf{Distributions of final orbital elements among planet-planet scattering simulations for which the hypothetical TOI-201~e was ejected.} Dashed black lines indicate the observed orbital elements (Table \ref{tab:pyttv_final}).  (\textbf{A}) Distribution of simulated $e_b$ values showing the present-day value is at the tail end. (\textbf{B}) Distribution of simulated $e_c$ values showing the present-day value is at the far end of the distribution. (\textbf{C}) Distribution of simulated mutual inclinations. (\textbf{D}) Distribution of simulated $a_c$.}
    \label{fig:scattering-hist}
\end{figure}

In 30\% of the simulations, the  hypothetical TOI-201~e is ejected and the eccentricities of TOI-201~b and c are excited. Among these simulations, $\sim 0.5 \%$ attained a final eccentricity of TOI-201~c of at least $0.6$ (Fig.~\ref{fig:scattering-hist}). One such example is depicted in Fig.~\ref{fig:scattering-example}. Notably, we find that the timescale for which the hypothetical TOI-201~e is ejected is typically $1-10$ kyr, which is several orders of magnitude less than the inferred age of the system. Therefore, due to the rapid instability timescale and relative paucity of high-eccentricity TOI-201~c's in our simulations, we find that planet-planet scattering may be plausible but requires a narrow range of initial conditions to replicate the system's observed architecture. Future studies could better evaluate the feasibility of this scenario by testing a wider range of initial conditions, which may lead to longer instability timescales or a greater efficiency of exciting TOI-201~c's eccentricity.

\subsubsection*{Possible Stellar Companions}
\begin{figure}
\centering
\includegraphics[width=\textwidth]{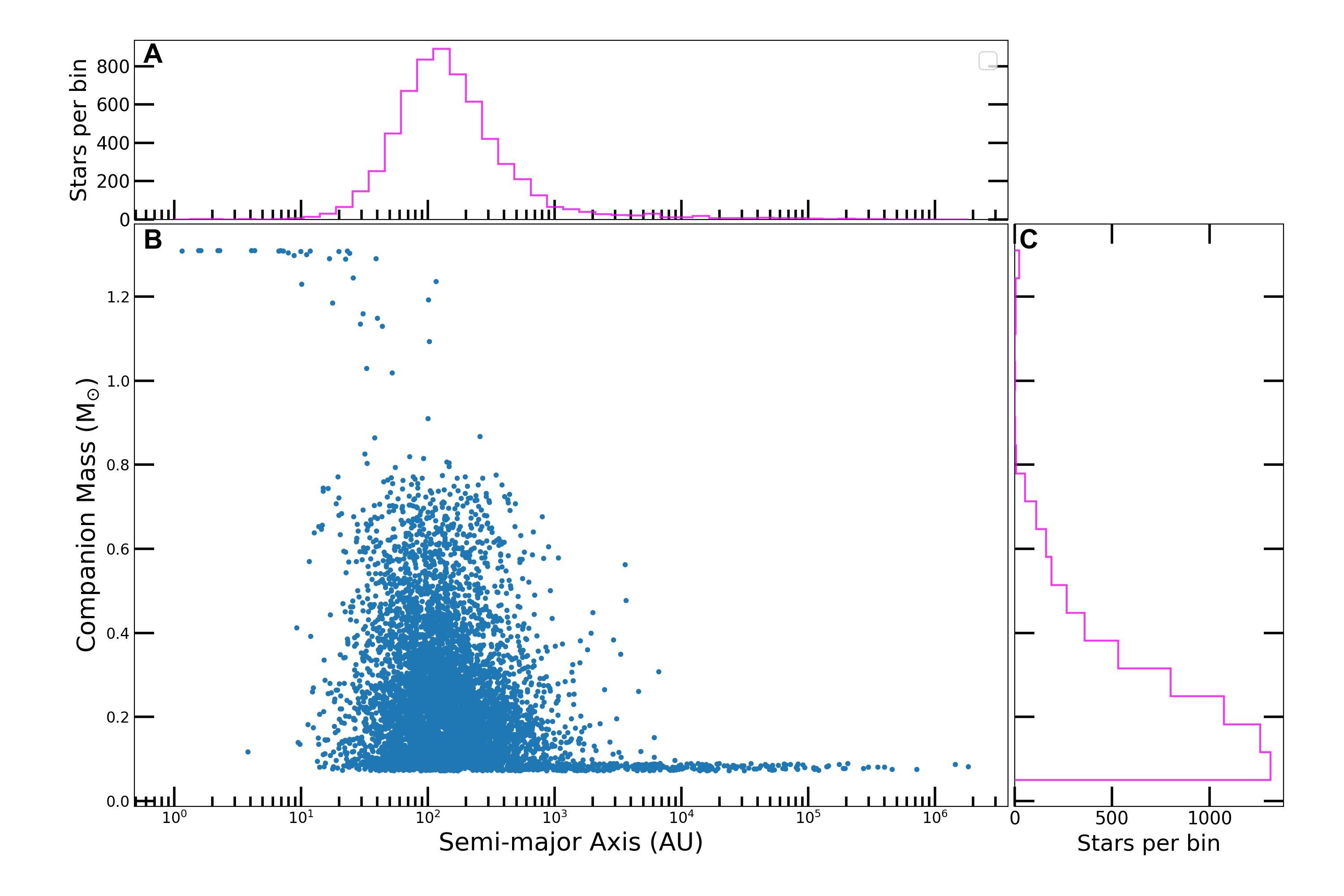}
\caption{\textbf{Stellar companions allowed by the existing data as determined using MOLUSC.} (\textbf{A}) Distribution of semi-major axes showing a peak at 100 AU. (\textbf{B}) Scatter plot of semi-major axis and companion masses for the allowed companions. (\textbf{C}) Distribution of companion masses showing a strong skew towards low masses.}
\label{fig:molusc}
\end{figure}

If the TOI-201 system is undergoing vZLK cycles due to a second star in the system, then this additional stellar companion would have been missed by existing observations. In order to quantify any possible undetected stellar companions, we use Multi-Observational Limits on Unseen Stellar Companions (MOLUSC) to generate a sample of potential companions consistent with the combination of the high-resolution imaging, RV data, Gaia astrometry (in the form of the RUWE), and Gaia imaging \cite{2021AJ....162..128W}. Of the 100,000 objects we generated, 20\% are consistent with the existing observations. The vast majority of these objects have masses less than 0.8 $M_{\odot}$, and most objects more massive than that have semi-major axes that would disrupt the orbits of the planets (see Fig. \ref{fig:molusc}). For the remaining stellar mass objects, the semi-major axes range from approximately 10 to 1000 AU.

\subsubsection*{Similar Systems}

There are currently three other systems with a close-in giant planet and a distant outer brown dwarf: Kepler-448 \cite{2017AJ....154...64M}, WASP-53 \cite{triaud2017}, and WASP-81 \cite{triaud2017}. Kepler-448 is particularly intriguing due the similarities between the giant planets in that system and TOI-201. Both warm Jupiters in the Kepler-448 and TOI-201 systems are moderately eccentric ($e \sim 0.3$) and show transit timing variations due to the outer companion. Both outer companions are highly eccentric ($e \sim 0.6-0.65$) and have similar masses and periods (Kepler-448 c has a mass of 22 Jupiter masses and period of 2500 days). Both systems show a significant nonzero mutual inclination between the giant planets, with a mutual inclination of $20^\circ$ in the Kepler-448 system compared to the $13^\circ$ between TOI-201~b and c. As with TOI-201, the mechanisms invoked to explain the architecture of Kepler-448 suffer from issues regarding fine-tuning.

Both WASP-53 and WASP-81 contain close-in inner giants, with both inner giants in the systems having orbital periods under 10 days. They do not exhibit detectable TTVs like TOI-201 and Kepler-448, however. Unlike in the case of TOI-201, high-eccentricity migration pathways could explain the WASP-53 and WASP-81 systems, as they both lack a second close-in planet that has to be preserved. 

Given how unlikely it is for a single stellar flyby to explain any individual system, it is even more unlikely that all of these systems underwent this process, given the fine-tuning required. In the case of planet-planet scattering, systems can and do form multiple giant planets, often in close proximity to one another. If this occurred, these giant planets were close enough to each other that one was scattered from the system, causing the outer companion to become eccentric. Nonetheless, this mechanism also suffers from a fine-tuning problem.

There is also a potential fifth system, WASP-132, but more observations are needed to determine if there is indeed a brown dwarf present \cite{2025A&A...693A.144G}.
WASP-132 contains a small inner planet, a hot Jupiter, an outer giant planet at 2.7 AU, and a long-term RV trend which can be a BD or a stellar companion. If the brown dwarf does exist, then the larger separations between planets could mean scattering never took place, and all of the giant planets that originally formed in the system were preserved. Conversely, the potential brown dwarf in the WASP-132 system could have driven or still be driving vZLK oscillations that sculpted the system's architecture.

Of these systems, TOI-201 has provided the most insight and is poised to continue to do so. It is the only system where the brown dwarf transits, which allows us to measure its radius and opens up opportunities for future observations to study its atmosphere. TOI-201 is also the only system with a confirmed brown dwarf and a super-Earth. It is significantly brighter than the other systems with a J-band magnitude that is 2 magnitudes brighter than the next brightest star. This makes it the target best suited for atmospheric characterization with the James Webb Space Telescope (JWST), as it is 7 times brighter than the other systems in the infrared region of the spectrum that JWST observes in.

\subsubsection*{Age–radius evolution of the brown dwarf companion}
TOI-201~c is one of five confirmed transiting companions at the canonical deuterium fusing mass threshold between giant planets and brown dwarfs of roughly 13 $M_J$\cite{baraffe2002, burrows2011, 2011ApJ...727...57S}, with the others being HATS-70 b \cite{zhou2019}, TOI-4603 b \cite{khandelwal2023}, TOI-4987 b \cite{subjak2024}, and TIC 4672985 b \cite{jones2024}. As with any transiting system, we have the opportunity to examine the radius evolution of the companion given the precise and accurate (when orbiting Sun-like main sequence stars) radius measurements. This is especially important in the case of transiting brown dwarfs as we predict that the radius monotonically contracts with age \cite{phillips2020, marley21, morley2024}, meaning that, for isolated brown dwarfs, younger objects will have larger radii than older objects of the same mass. In the case of most transiting brown dwarfs, we must consider the effects of the host star in its energy contribution to the atmosphere of the brown dwarf via irradiation and tidal heating. However, it is clear from the large scaled semi-major axis of ($a/R_\star=716$) and average incident flux ($\langle F \rangle=0.15\, S_{\oplus}$) received by TOI-201~c that we can treat this low-mass brown dwarf as an isolated object with regard to its radius evolution.

\begin{figure}
\centering
\includegraphics[width=\textwidth]{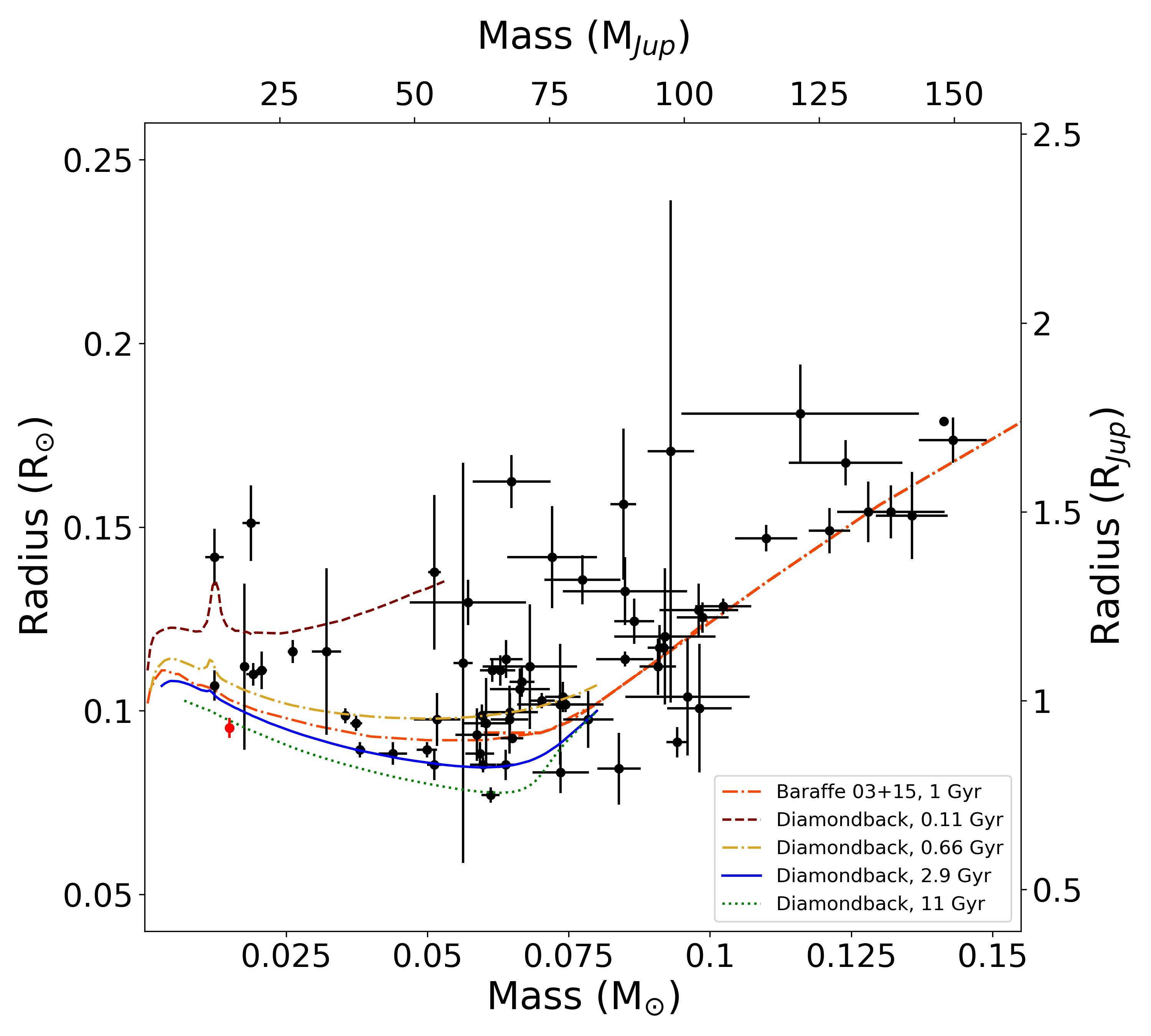}
\caption{\textbf{Mass-radius diagram from transiting brown dwarfs and low-mass stars.} The masses and radii are taken from \cite{2024MNRAS.533.2823H} and references therein. The models are for solar metallicity brown dwarfs and low-mass stars of different ages \cite{2003A&A...402..701B, 2015A&A...577A..42B, morley2024}. TOI-201~c is denoted by the red circle and lies well below where predicted for the system's age of 0.66 Gyr.}\label{fig:massrad}
\end{figure}

This makes TOI-201~c an excellent benchmark system for age–radius evolutionary models for substellar objects. Fig. \ref{fig:massrad} indicates how TOI-201~c’s mass, radius, and age compare to brown dwarf and low-mass star models at solar metallicity \cite{2003A&A...402..701B, 2015A&A...577A..42B, phillips2020}. Interestingly, it lies below where models predict based on the host star's age of approximately 666 Myr. However, it is important to note that there can be significant scatter in brown dwarf models, due to degeneracies between model parameters, including metallicity and the presence, or lack thereof, of clouds \cite{burrows2011, 2021ApJ...921...95Z}. When we attempt to consider parameters like metallicity, we find that the models better approximate the radius of the brown dwarf when we assume a metal-poor, cloud-free atmosphere. Even then, the radius of the brown dwarf lies 1.9-$\sigma$ below the most generous interpretation of the system (i.e. assuming the oldest plausible system age of 2.9 Gyr at the lowest metallicity the models account for). 
Atmospheric characterization of the brown dwarf would constrain these parameters and help determine why it may be truly  smaller than predicted by existing models.

\subsubsection*{Future Prospects}

The dynamical history could be further constrained with a measurement of the system's obliquity relative to the spin axis of the star. While the mutual inclinations between the three known bodies are constrained in this work, the overall obliquity of the system is not. A Rossiter-McLaughlin measurement of TOI-201~b is the most practical way to achieve this. There are multiple transits through the end of 2026 that are observable from sites in Chile and Australia with the capabilities of measuring the expected 30 m/s signal.

A more precise mass measurement of TOI-201~d would reveal whether it is a suitable target for atmospheric characterization with the James Webb Space Telescope (JWST). While its size together with its proximity to the host star likely places it in the rocky planet regime and unlikely to have an atmosphere \cite{Owen2013,Rogers2015,Fulton2018}, it could be a candidate for secondary eclipse spectroscopy to probe the composition and features of its surface. TOI-201~b is a much more promising candidate for transmission spectroscopy, having a Transmission Spectroscopy Metric (TSM) value of 110 which places it in the first quartile for prioritization among giant exoplanets \cite{Kempton2018}. Since TOI-201~b is expected to have formed at least one Hill radius interior to TOI-201~c's orbit (i.e. with an initial semi-major axis $<$3.7 AU), its atmospheric metallicity is predicted to be super-stellar since the planet would likely not have formed in the outer, gas-rich region of the protoplanetary disk \cite{Feinstein2025}. A measurement of its atmospheric metallicity could test this hypothesis.

Observing a full transit of TOI-201~c will help refine its orbital parameters, specifically its orbital period, and resolve the degeneracy between the transit impact parameter and duration. Additional RV measurements in the next several years, especially in the months preceding the next transit, will help reduce the uncertainty on the transit timing. Afterward, a transit observation can be executed using a combination of ground-based telescopes across the world, including from citizen scientists, as has been done with other single-transit planets \cite{2025AJ....170...41E}. Despite these challenges, the brown dwarf is a very promising target for atmospheric characterization. Its long orbital period makes it fairly isolated from its host star and, as a result, should not be inflated like many known transiting brown dwarfs. This makes TOI-201~c an important benchmark for understanding the structure of brown dwarfs, as well as understanding how their radii evolve over time. Atmospheric characterization could reveal important information about its metallicity, which in turn would inform how and where it formed. The existence of a brown dwarf desert near 45 Jupiter masses and differences in the properties between brown dwarfs above and below the desert suggest that high-mass and low-mass brown dwarfs form in different ways \cite{2014MNRAS.439.2781M}. High-mass brown dwarfs likely form similarly to stars, through fragmentation of the molecular cloud while low-mass brown dwarfs like TOI-201~c are thought to form similarly to planets, through either core accretion or gravitational instability. While it is too close to have formed in-situ through gravitational instability, it is also more massive than what models typically predict can form through core accretion \cite{2018ApJ...853...37S}, although the host star's high metallicity could make it possible \cite{2017A&A...602A..38M}.

\clearpage





\begin{figure*}[t!]
\centering
\includegraphics[width=\textwidth]{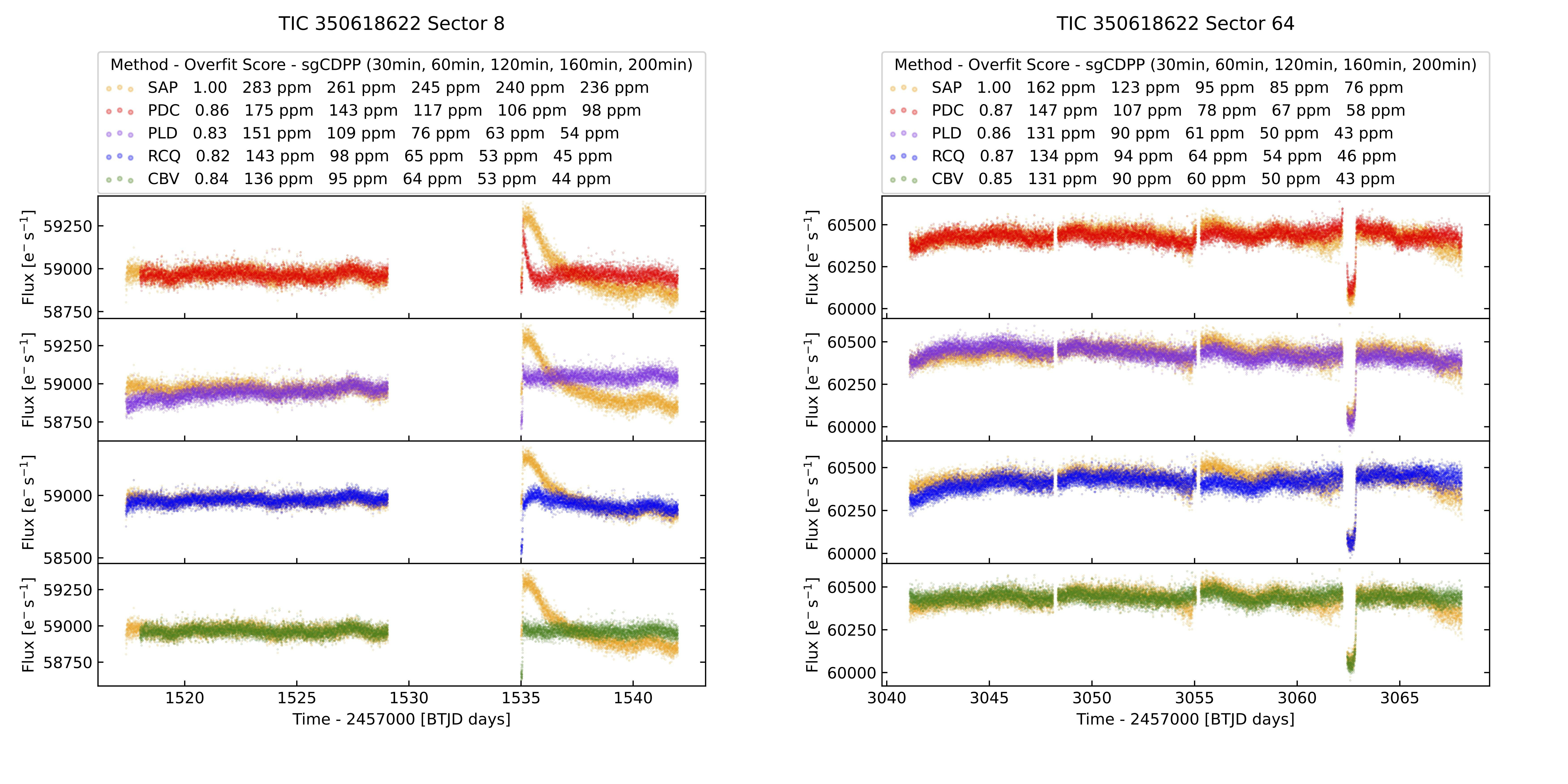}
\caption{\textbf{2-minute TESS SAP light curves plotted against different corrected light curves for Sectors 8 and 64.} The Pixel Level Decorrelation (PLD), Regression Corrector with Quaternions (RCQ), and Cotrending Basis Vectors (CBV) corrected light curves do not exhibit the ramp-like features seen in the PDCSAP light curve. 
}
\label{fig:spoc_lc_corrections}
\end{figure*}

\begin{figure*}[t!]
\centering
\includegraphics[width=\textwidth]{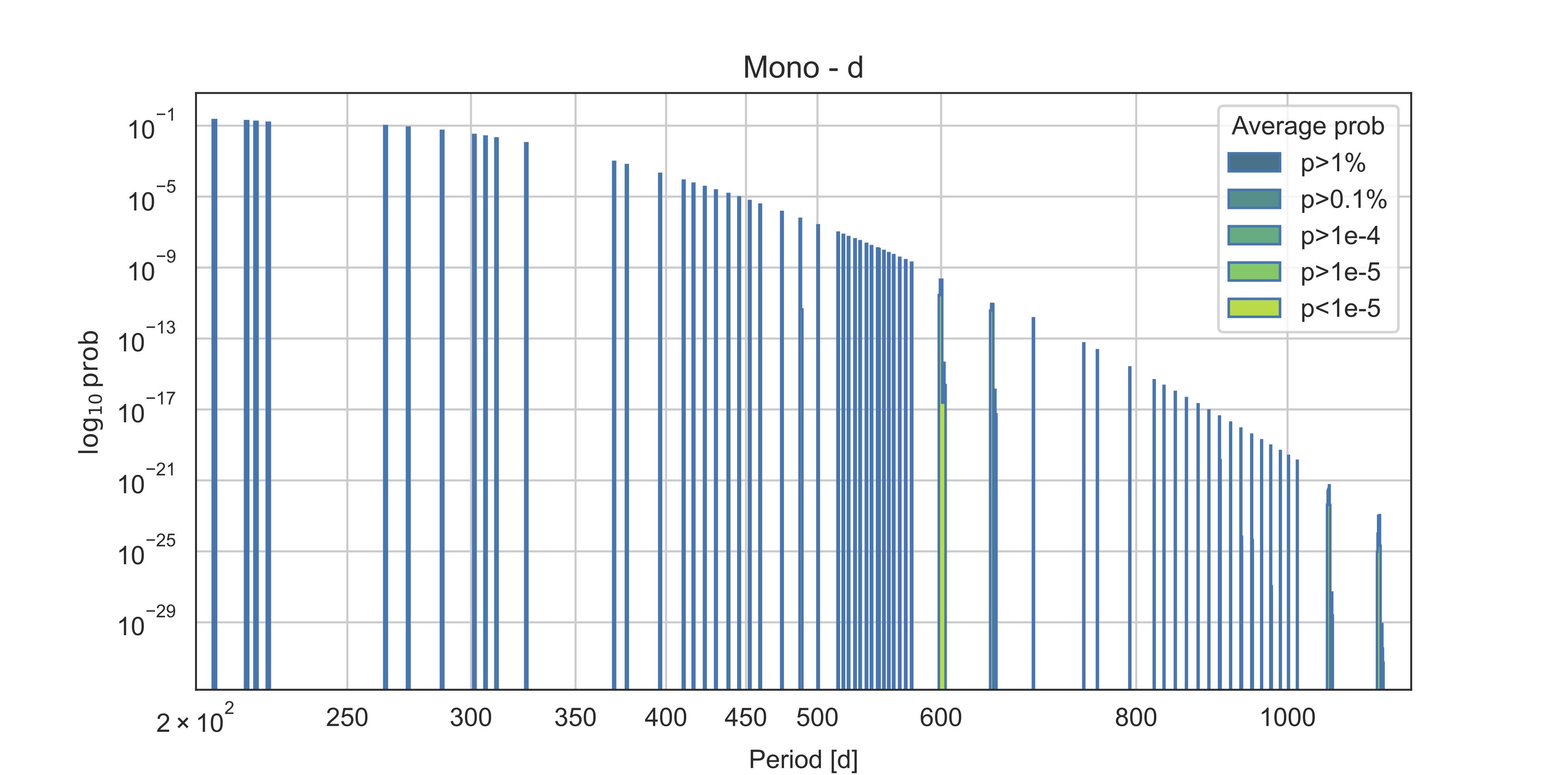}
\caption{\textbf{Periods allowed for TOI-201~c based on TESS photometry alone as determined by \texttt{MonoTools}.} Periods below 500 days have uncertainties on the order of a few hours, making them testable from the ground. The marginal probabilities decrease with increasing period as expected from the geometric transit probability.
}
\label{fig:monotools}
\end{figure*}

\begin{figure*}[t!]
\centering
\includegraphics[width=\textwidth]{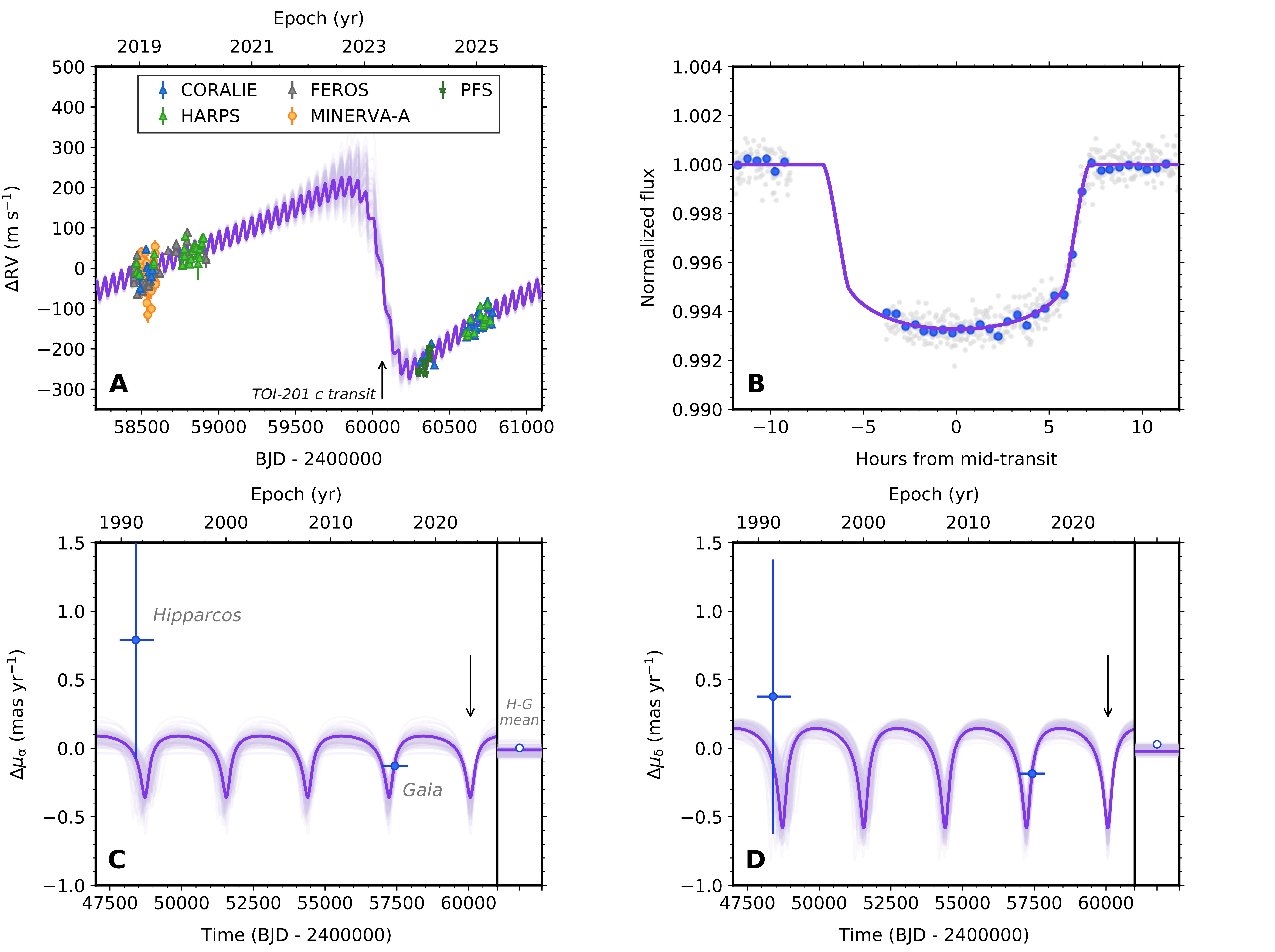}
\caption{\textbf{Joint model of RVs, transits, and \textit{Hipparcos-Gaia} astrometry.}
(\textbf{A}) RVs and best-fit two-planet RV model. (\textbf{B}) TESS photometry and best-fit transit of model TOI-201~c. (\textbf{C}) Proper motion in right ascension and best-fit astrometric model for TOI-201~c. The three astrometric points are labelled; the \textit{Hipparcos-Gaia} mean proper motion is shown in the side panel (see \cite{Venner2021} for further details on the format). (\textbf{D}) Proper motion in declination and best-fit astrometric model, formatted as in (\textbf{C}). In panels (\textbf{A}, \textbf{C}, \textbf{D}), the epoch of the observed transit of TOI-201~c is marked by an arrow. The \textit{Gaia} observations are coincident with the preceding 2015 periastron passage of TOI-201~c, which explains the significant astrometric acceleration.
}
\label{fig:joint_astrometry}
\end{figure*}

\begin{figure}
\centering
\includegraphics[width=\textwidth]{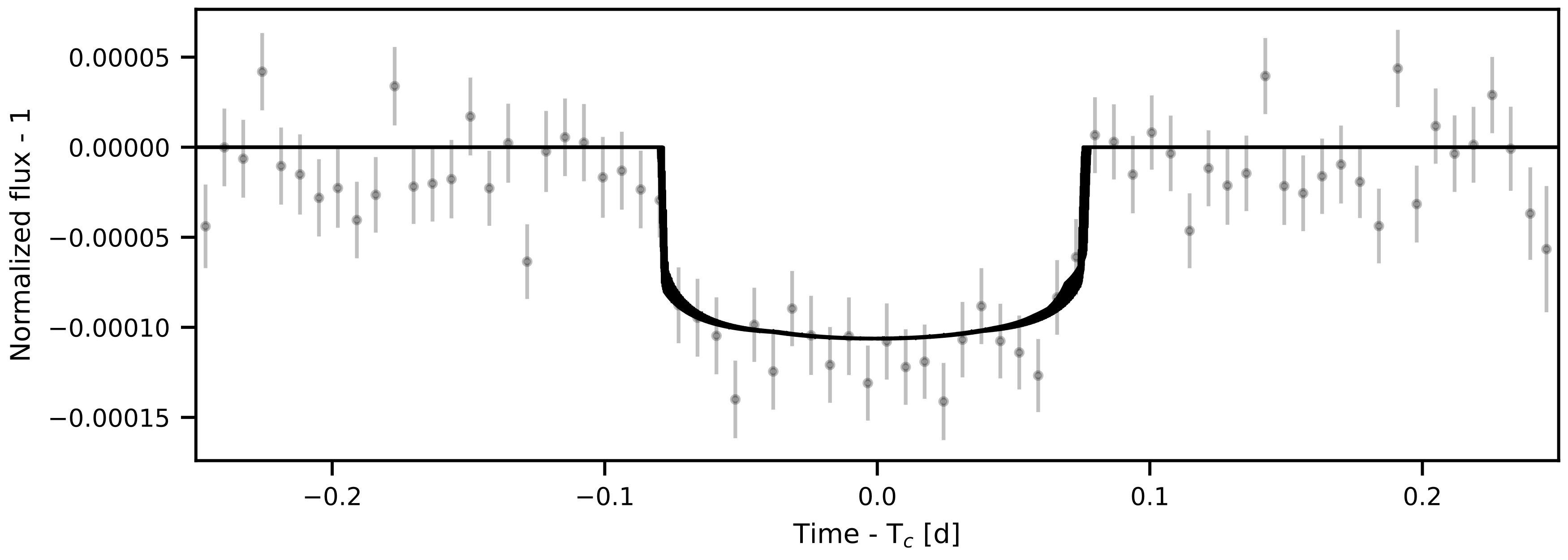}
\caption{\textbf{Phase-folded transits of TOI-201~d in the photodynamical model.} The TESS data is binned for clarity.}
\label{fig:inditransit_201c}
\end{figure}

\begin{figure}
\centering
\includegraphics[width=\textwidth]{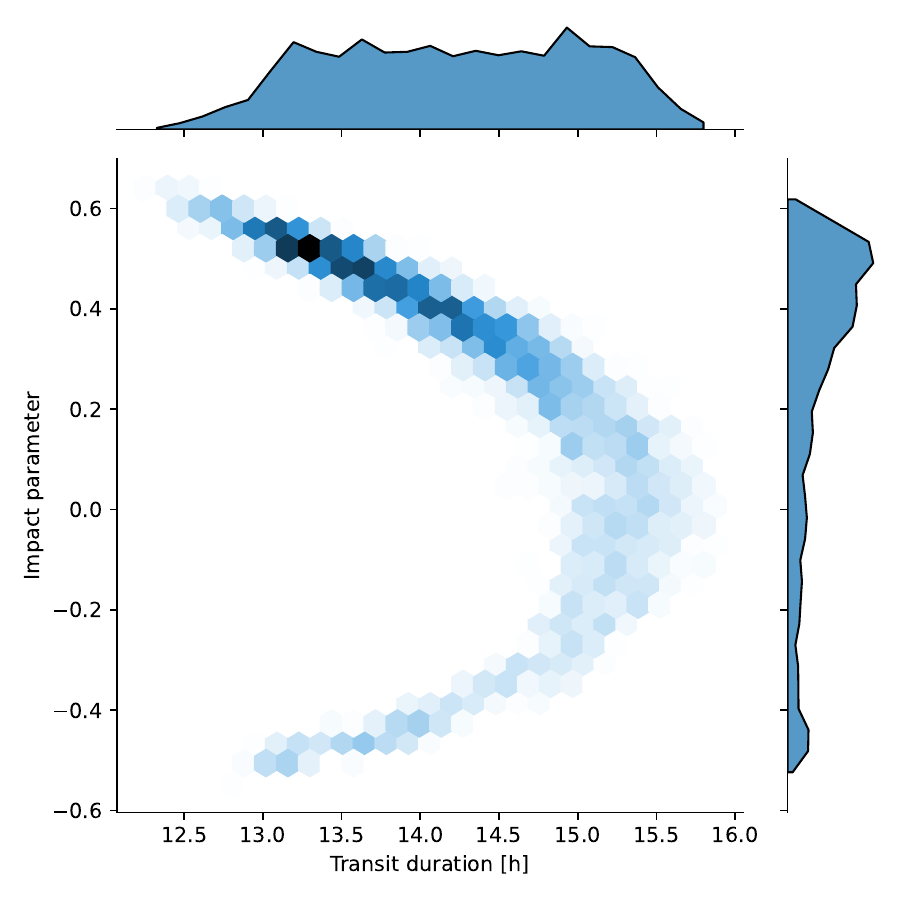}
\caption{\textbf{Duration vs impact parameter for TOI-201~c.} The degeneracy between the two parameters is apparent and is due to the transit not being observed in full.}
\label{fig:duration}
\end{figure}

\begin{figure}
    \centering
    \includegraphics[width=\linewidth]{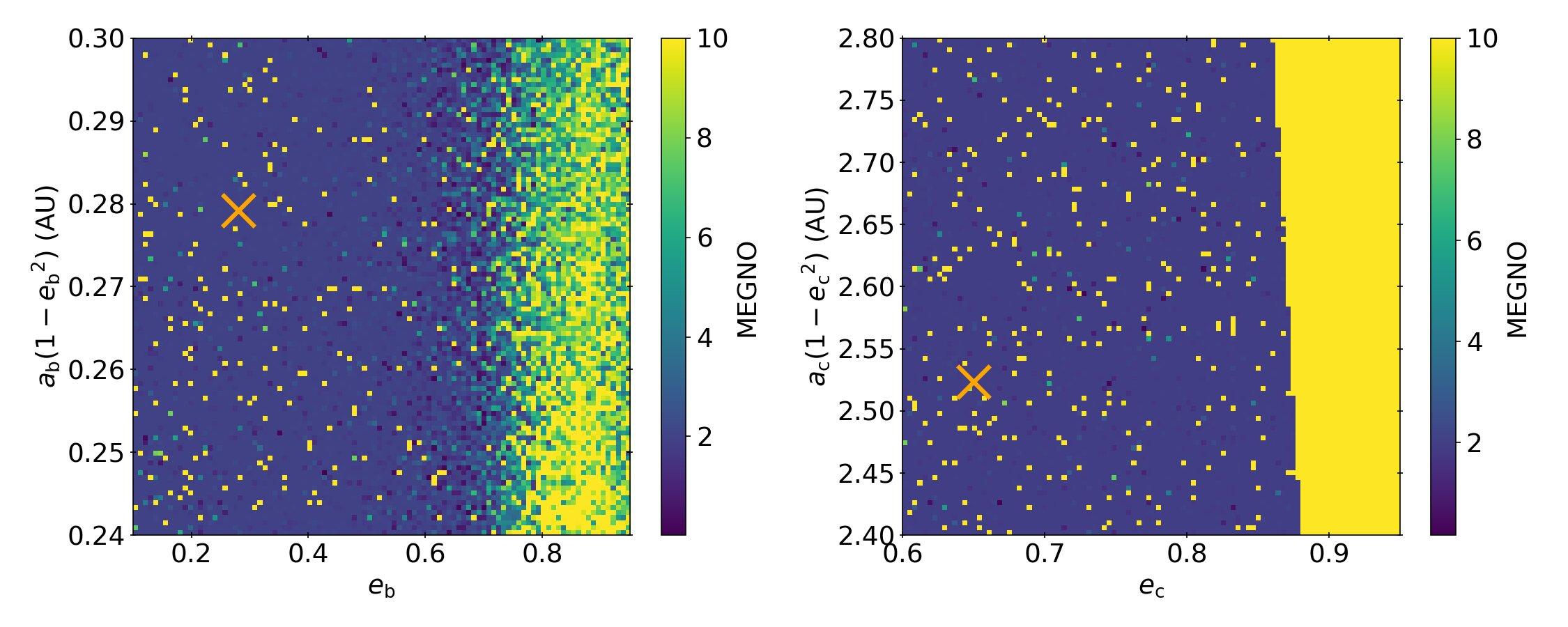}
    \caption{\textbf{MEGNO map for TOI-201~b (left) and TOI-201~c (right) for a variety of angular momenta and eccentricities.} The orange marker on each map corresponds to the observed orbital parameters of the system. Since regions where $e \gtrsim 0.9$ result in chaotic evolution and $a(1-e^2)$ is conserved during tidal interactions, we deduce that the system is not consistent with a history of high-eccentricity migration.}
    \label{fig:high-e-megno}
\end{figure}

\begin{table}[h!]
\renewcommand{\arraystretch}{0.8}
\centering
\caption{\textbf{Observed and derived stellar parameters for TOI-201.} The parameters from this work were obtained from an isochrone fit.}
\label{tab:star_parameters}
\begin{tabular}{lll}
\hline
Parameter   & Value & Source\\
\hline
TIC & 350618622 & TICv8 \cite{stassun2018} \\
Additional Identifiers & HD 39474; HIP 27515 & - \\
Right Ascension & 05:49:36.43 & Gaia DR3 \cite{gaia2023} \\
Declination & -54:54:37.49 & Gaia DR3 \\
$\mu_{\alpha}$ (mas yr$^{-1}$) & $8.032\pm0.018$ & Gaia DR3 \\
$\mu_{\delta}$ (mas yr$^{-1}$) & $66.633\pm0.019$ & Gaia DR3 \\
Parallax (mas) & $8.9141 \pm 0.0142$ & Gaia DR3 \\
$B$ (mag) & $9.793 \pm 0.29$ & APASS DR10 \cite{henden2019} \\
$V$ (mag) & $9.036 \pm 0.020$ & APASS DR10 \\
$G$ (mag) & $8.949289 \pm 0.002761$ & Gaia DR3 \\
$B_P$ (mag) & $9.203999 \pm 0.002826$ & Gaia DR3 \\
$R_P$ (mag) & $8.532517 \pm 0.003806$ & Gaia DR3 \\
$TESS$ (mag) & $8.582 \pm 0.006$ & TICv8 \\
$J$ (mag) & $8.103 \pm 0.029$ & 2MASS \cite{cutri2003} \\
$H$ (mag) & $7.923 \pm 0.036$ & 2MASS \\
$K_S$ (mag) & $7.846 \pm 0.024$ & 2MASS \\
$W_1$ (mag) & $7.782 \pm 0.028$ & TICv8 \\
$W_2$ (mag) & $7.829 \pm 0.020$ & TICv8 \\
$W_3$ (mag) & $7.834 \pm 0.016$ & TICv8 \\
$W_4$ (mag) & $7.691 \pm 0.092$ & TICv8 \\
$M_{\star}\, (M_{\odot})$ & $1.32^{+0.02}_{-0.04}$ & This work \\
$R_{\star}\, (R_{\odot})$ & $1.31 \pm 0.01$ & This work \\
$\log g$ ($\log$ cm~s$^{-2}$) & $4.33^{+0.01}_{-0.02}$& This work \\
$L_{\star}\, (L_{\odot})$ & $2.61 \pm 0.12$ & This work \\
$[\mathrm{Fe/H}]$ & $0.19 \pm 0.06$ & This work \\
$T_{\rm{eff}}$ (K) & $6423^{-90}_{+86}$ & This work \\
Age (Myr) & $666^{+673}_{-442}$ & This work \\
\hline
\end{tabular}
\end{table}

\begin{table}[h!]
\caption{\textbf{Orbital parameters for TOI-201~b and c from our joint model incorporating RVs, transits, and astrometry.} The parameters shown here are consistent with the values we obtain with the full photodynamical model.}
\centering
\begin{tabular}{lcc}
\hline
Parameter & TOI-201~b & TOI-201~c \\
\hline
~~Orbital period $P$ (d) & $52.979$ (fixed) & $2834^{+43}_{-38}$ \\
~~Orbital period $P$ (yr) & $0.145$ (fixed) & $7.76^{+0.12}_{-0.10}$ \\
~~Semi-amplitude $K$ (m~s$^{-1}$) & $23.8\pm1.4$ & $230^{+26}_{-19}$ \\
~~Eccentricity $e$ & $0.287^{+0.038}_{-0.036}$ & $0.610^{+0.047}_{-0.041}$ \\
~~Argument of periastron $\omega$ (deg) & $89.3^{+7.5}_{-7.4}$ & $98.0\pm6.2$ \\
~~$T_\text{p}$ (BJD) & $2459965.3\pm0.6$ & $2460074.5^{+12.9}_{-9.7}$ \\
\hline
~~$T_0$ (BJD) & $2459965.3944\pm0.0008$ & $2460062.579^{+0.016}_{-0.021}$ \\
~~Impact parameter $b$ & $0.73\pm0.03$ & $0.41^{+0.13}_{-0.22}$ \\
~~Radius ratio $R_\text{p}/R_*$ & $0.07928\pm0.0009$ & $0.07736\pm0.0011$ \\
\hline
~~Orbital inclination $i$ (deg) & $88.82\pm0.04$ & $89.916^{+0.044}_{-0.022}$ \\
~~Longitude of node $\Omega$ (deg) & -- & $212\pm11$ \\
\hline
~~Mass $M_\text{p}$ ($M_J$) & $0.505\pm0.031$ & $15.4^{+1.0}_{-0.8}$ \\
~~Radius $R_\text{p}$ ($R_J$) & $1.030\pm0.023$ & $1.005\pm0.023$ \\
\hline
\end{tabular}
\label{tab:joint_astrometry}
\end{table}

\begin{table}[h!]
\renewcommand{\arraystretch}{0.55}
    \centering
    \caption{\textbf{Photodynamical model parameters with their priors and posteriors.} $\mathcal{N}(\mu, \sigma)$ denotes a normal prior with a mean $\mu$ and standard deviation $\sigma$, and $\mathcal{U}(a,b)$ denotes a uniform distribution from $a$ to $b$. The osculating orbital elements are valid for the reference time $T_{\mathrm{ref}}=2458376.052$.}
    \begin{tabular}{llrr}
    \hline
    Model Parameter & Unit & Prior & Posterior  \\
    \hline
    \textit{Stellar parameters} & & & \\
    \hline
    $R_\star$ & $R_\odot$ & $\mathcal{N}(1.32,0.01)$ & $1.31\pm0.01$ \\
    $M_\star$ & $M_\odot$ & $\mathcal{N}(1.32,0.03)$ & $1.33\pm0.03$ \\
    ${q}_1$ & & $\mathcal{U}(0,1)$ & $0.23\pm0.03$ \\
    ${q}_2$ & & $\mathcal{U}(0,1)$ & $<0.36$\\ 
    \\
    \textit{RV parameters} & & & \\
    \hline
    $\gamma_{\mathrm{MINERVA}}$ & m\,$\mathrm{s}^{-1}$ & $\mathcal{N}(0,50)$ & $-3\pm5$\\
    $\gamma_{\mathrm{HARPS}}$ & m\,$\mathrm{s}^{-1}$ & $\mathcal{N}(16700,50)$ & $16731\pm2$\\
    $\gamma_{\mathrm{CORALIE}}$ & m\,$\mathrm{s}^{-1}$ & $\mathcal{N}(16700,50)$ & $16846\pm3$\\
    $\gamma_{\mathrm{FEROS}}$ & m\,$\mathrm{s}^{-1}$ & $\mathcal{N}(16865,50)$ & $16866\pm4$\\
    $\gamma_{\mathrm{PFS}}$ & m\,$\mathrm{s}^{-1}$ & $\mathcal{N}(250,50)$ & $232\pm4$\\
    $\log_{10}\sigma_{\mathrm{MINERVA}}$ & $\log_{10}$ m\,$\mathrm{s}^{-1}$ & $\mathcal{N}(-1,1)$ & $1.53\pm0.05$\\
    $\log_{10}\sigma_{\mathrm{HARPS}}$ & $\log_{10}$ m\,$\mathrm{s}^{-1}$ & $\mathcal{N}(-3,0.1)$ & $-3.0\pm0.1$\\
    $\log_{10}\sigma_{\mathrm{CORALIE}}$ & $\log_{10}$ m\,$\mathrm{s}^{-1}$ & $\mathcal{N}(-1,1)$ & $-0.9\pm0.8$\\
    $\log_{10}\sigma_{\mathrm{FEROS}}$ & $\log_{10}$ m\,$\mathrm{s}^{-1}$ & $\mathcal{N}(-1,1)$ & $1.28\pm0.05$\\
    $\log_{10}\sigma_{\mathrm{PFS}}$ & $\log_{10}$ m\,$\mathrm{s}^{-1}$ & $\mathcal{N}(-1,1)$ & $0.82\pm0.09$\\
    \\
    \textit{TOI-201~d} & & & \\
    \hline
    $P_{d}$ & days & $\mathcal{N}(5.849,0.005)$ & $5.8489\pm0.0001$ \\
    $T_{\mathrm{0,d}}$ & BJD & $\mathcal{N}(2458374.033, 0.003)$ & $2458374.032\pm0.003$ \\
    $\log_{10}\, M_\mathrm{p,d}$ &$\log_{10}\,M_\odot$ & $\mathcal{N}(-5.07,0.15)$ & $-4.8\pm0.1$ \\         
    $R_\mathrm{p,d}/R_\star$ & & $\mathcal{N}(0.01,0.001)$ & $0.0097\pm0.0005$ \\ 
    $\sqrt{e_\mathrm{d}}\cos{\omega_\mathrm{d}}$ & & $\mathcal{U}(-1, 1)$ & $-0.06\pm0.09$ \\ 
    $\sqrt{e_\mathrm{d}}\sin{\omega_\mathrm{d}}$ & & $\mathcal{U}(-1, 1)$ & $0.47\pm0.09$\\      
    $b_\mathrm{d}$ & & $\mathcal{U}(-1, 1)$ & $-0.1\pm 0.3$ \\
    $\Omega_\mathrm{d}$ & rad & $\mathcal{U}(0.5\pi,1.5\pi)$ & $3.1\pm0.4$\\
    \\
    \textit{TOI-201~b} & & \\
    \hline
     $P_\mathrm{b}$ & days & $\mathcal{N}(52.980,0.005)$ & $52.9786\pm0.0001$ \\
    $T_{\mathrm{0,b}}$ & BJD & $\mathcal{N}(2458376.0520, 0.0002)$ & $2458376.0521\pm0.0002$ \\
    $\log_{10}\, M_\mathrm{p,b}$ &$\log_{10}\,M_\odot$ & $\mathcal{U}(-4.0, -2.9)$ & $-3.31\pm0.01$ \\        
    $R_\mathrm{p,b}/R_\star$ & & $\mathcal{N}(0.07,0.003)$ & $0.0798\pm0.0004$ \\ 
    $\sqrt{e_\mathrm{b}}\cos{\omega_\mathrm{b}}$ & & $\mathcal{U}(-1, 1)$ & $0.522\pm0.008$ \\ 
    $\sqrt{e_\mathrm{b}}\sin{\omega_\mathrm{b}}$ & & $\mathcal{U}(-1, 1)$ & $0.06\pm0.02$\\      
    $b_\mathrm{b}$ & & $\mathcal{U}(-1, 1)$ & $-0.741\pm 0.006$ \\
    $\Omega_\mathrm{b}$ & rad & $\mathcal{U}(0.5\pi,1.5\pi)$ & $3.5\pm 0.2$\\
    \\
    \textit{TOI-201~c} & & \\
    \hline
    $P_\mathrm{c}$ & days & $\mathcal{U}(2300,3500)$ & $2890\pm20$ \\
    $T_{\mathrm{0,c}}$ & BJD & $\mathcal{N}(2460062.6, 0.1)$ & $2460062.59\pm0.02$\\
    $\log_{10}\,M_\mathrm{p,c}$ &$\log_{10}\,M_\odot$ & $\mathcal{U}(-2.5, -1.2)$ & $-1.824\pm0.009$\\         
    $R_\mathrm{p,c}/R_\star$ & & $\mathcal{N}(0.07,0.003)$ & $0.073\pm0.002$\\ 
    $\sqrt{e_\mathrm{c}}\cos{\omega_\mathrm{c}}$ & & $\mathcal{U}(-1, 1)$ & $0.800\pm0.0005$\\ 
    $\sqrt{e_\mathrm{c}}\sin{\omega_\mathrm{c}}$ & & $\mathcal{U}(-1, 1)$ & $-0.09\pm0.03$\\      
    $b_\mathrm{c}$ & & $\mathcal{U}(-1, 1)$ & $0.4\pm 0.3$ \\
    $\Omega_\mathrm{c}$ & rad & $\mathcal{N}(3.70, 0.19)$ & $3.7\pm 0.2$\\
    \hline
    \hline
    \end{tabular}
    \label{tab:pyttv_fitting}
\end{table}

\begin{table}[h!]
\renewcommand{\arraystretch}{0.7}
    \centering
    \caption{\textbf{Photodynamical final posterior parameters.} The osculating orbital elements are valid for the reference time $T_{\mathrm{ref}}=2458376.052$.}
    \begin{tabular}{llrrr}
    \hline
    \hline
    Parameter & Unit & TOI-201~d& TOI-201~b &  TOI-201~c\\
    \hline
    $P$ & days & $5.8489\pm0.0001$ & $52.9786\pm0.0001$ & $2890\pm20$\\
    $T_0$ & BJD-2458000 & $374.032\pm0.003$ & $376.0521\pm0.0002$ & $2062.59\pm0.02$\\
    $M_\mathrm{p}$ & $M_\oplus$ &  $5.8\pm2$ & $164\pm5$ & $4990\pm100$ \\  
    $R_\mathrm{p}$ & $R_\oplus$ &  $1.39\pm0.07$ & $11.4\pm0.1$ & $10.4\pm0.3$\\ 
    $\rho_\mathrm{p}$ & g\,$\mathrm{cm}^{-3}$ & $11\pm4$ & $0.61\pm0.02$ & $24\pm2$\\
    $\mathrm{T_{14}}$ & h & $3.76\pm0.08$ & $4.13\pm0.01$ & $13.1\pm0.8$\\
    $e$ & & $0.3\pm0.1$ & $0.275\pm0.009$ & $0.651\pm0.006$\\
    $\omega$ & $^\circ$ & $340\pm80$ & $83\pm2$ & $96\pm2$\\ 
    $i$ & $^\circ$ & $91.7\pm1.6$ & $91.18\pm0.02$ & $89.92^{+0.10}_{-0.04}$\\ 
    $\Omega$ & $^\circ$ & $175\pm20$ & $198\pm10$ & $211\pm11$\\ 
    $a/R_\star$ & & $11.4\pm0.1$ & $49.7\pm0.4$ & $716\pm8$\\
    $a$ & AU & $0.0698\pm0.0005$ & $0.303\pm0.002$ & $4.37\pm0.04$\\
    \hline
    \hline
    \end{tabular}
    \label{tab:pyttv_final}
\end{table}

\begin{table}[]
\renewcommand{\arraystretch}{0.7}
    \centering
    \caption{\textbf{New time series radial velocities from CORALIE.}}
\begin{tabular}{llll}
    \hline
    \hline
    Time (BJD-2457000) & RV (m s$^{-1}$) & RV Uncertainty (m s$^{-1}$) & Instrument \\
    \hline
    3311.6364   & 16610.90  & 9.16        & CORALIE    \\
    3344.6467   & 16622.50  & 18.63       & CORALIE    \\
    3362.6982   & 16622.35 & 15.39       & CORALIE    \\
    3381.5481   & 16655.99 & 7.90         & CORALIE    \\
    3401.5122   & 16601.94 & 7.72        & CORALIE    \\
    3602.7809   & 16686.58 & 8.57        & CORALIE    \\
    3609.7397   & 16670.91 & 8.11        & CORALIE    \\
    3616.7343   & 16674.18 & 9.68        & CORALIE    \\
    3623.7228   & 16679.83 & 7.60         & CORALIE    \\
    3627.5930    & 16695.77 & 8.41        & CORALIE    \\
    3634.6395   & 16715.16 & 11.59       & CORALIE    \\
    3640.7489   & 16710.02 & 8.69        & CORALIE    \\
    3662.6176   & 16675.55 & 9.74        & CORALIE    \\
    3668.7606   & 16689.79 & 8.77        & CORALIE    \\
    3670.5819   & 16694.50  & 10.12       & CORALIE    \\
    3678.6534   & 16718.74 & 9.23        & CORALIE    \\
    3685.5400     & 16733.62 & 11.95       & CORALIE    \\
    3694.7396   & 16732.99 & 10.51       & CORALIE    \\
    3709.5339   & 16706.52 & 10.09       & CORALIE    \\
    3716.6576   & 16695.25 & 12.68       & CORALIE    \\
    3723.5855   & 16714.22 & 9.18        & CORALIE    \\
    3730.5526   & 16707.76 & 10.50        & CORALIE    \\
    3747.5246   & 16760.30  & 9.73        & CORALIE    \\
    3754.5644   & 16745.53 & 8.76        & CORALIE    \\
    3762.5604   & 16720.02 & 10.22       & CORALIE    \\
    3772.5339   & 16703.63 & 11.30        & CORALIE    \\
    3779.4784   & 16732.45 & 9.72        & CORALIE    \\
    \hline
    \hline
\label{tab:rv_values_coralie}
\end{tabular}
\end{table}

\begin{table}[]
\renewcommand{\arraystretch}{0.7}
    \centering
    \caption{\textbf{New time series radial velocities from HARPS.}}
\begin{tabular}{llll}
    \hline
    \hline
    Time (BJD-2457000) & RV (m s$^{-1}$) & RV Uncertainty (m s$^{-1}$) & Instrument \\
    \hline
    3603.6999   & 16569.70  & 9.25        & HARPS      \\
    3615.8744   & 16561.90  & 5.02        & HARPS      \\
    3618.8081   & 16562.23 & 6.71        & HARPS      \\
    3621.6992   & 16570.73 & 4.51        & HARPS      \\
    3635.8250    & 16604.25 & 5.87        & HARPS      \\
    3660.5947   & 16569.28 & 4.36        & HARPS      \\
    3699.7400     & 16636.26 & 5.21        & HARPS      \\
    3702.6655   & 16612.15 & 3.82        & HARPS      \\
    3720.5526   & 16588.90  & 4.93        & HARPS      \\
    3724.6358   & 16594.12 & 3.86        & HARPS      \\
    3731.5862   & 16603.57 & 3.83        & HARPS      \\
    3733.5648   & 16609.06 & 4.73        & HARPS      \\
    3747.6424   & 16641.06 & 4.25        & HARPS      \\
    3764.5931   & 16599.30  & 5.20         & HARPS      \\
    \hline
    \hline
\label{tab:rv_values_harps}
\end{tabular}
\end{table}

\begin{table}[]
\renewcommand{\arraystretch}{0.7}
    \centering
    \caption{\textbf{Time series radial velocities from PFS.}}
\begin{tabular}{llll}
    \hline
    \hline
    Time (BJD-2457000) & RV (m s$^{-1}$) & RV Uncertainty (m s$^{-1}$) & Instrument \\
    \hline
    3298.6407   & -28.42   & 2.56        & PFS        \\
    3298.7281   & -26.93   & 2.21        & PFS        \\
    3301.6846   & -18.07   & 2.51        & PFS        \\
    3301.7776   & -19.36   & 2.38        & PFS        \\
    3334.5868   & -0.30     & 1.84        & PFS        \\
    3334.6753   & 1.94     & 1.58        & PFS        \\
    3336.5880    & 0.32     & 1.94        & PFS        \\
    3336.6665   & 0.00        & 1.67        & PFS        \\
    3338.6077   & -12.25   & 1.58        & PFS        \\
    3338.6739   & -9.47    & 1.55        & PFS        \\
    3341.5710    & -29.83   & 1.53        & PFS        \\
    3341.6600     & -26.93   & 1.61        & PFS        \\
    3369.5364   & 27.97    & 2.00           & PFS        \\
    3369.5424   & 27.54    & 2.04        & PFS        \\
    3370.5238   & 14.63    & 2.07        & PFS        \\
    3370.5297   & 7.02     & 1.97        & PFS        \\
    3370.6154   & 15.11    & 2.04        & PFS        \\
    3372.5267   & 38.40     & 1.70        & PFS        \\
    3372.5978   & 33.18    & 1.75        & PFS \\
    \hline
    \hline
\label{tab:rv_values_pfs}
\end{tabular}
\end{table}


\clearpage 

\paragraph{Caption for Data S1.}
\textbf{CORALIE radial velocity time series.}
Machine readable file of CORALIE radial velocity time series as Barycentric Julian Dates (BJD), radial velocities, and associated uncertainties in m s$^{-1}$. 

\paragraph{Caption for Data S2.}
\textbf{FEROS radial velocity time series.}
Machine readable file of FEROS radial velocity time series as Barycentric Julian Dates (BJD), radial velocities, and associated uncertainties in m s$^{-1}$. 

\paragraph{Caption for Data S3.}
\textbf{HARPS radial velocity time series.}
Machine readable file of HARPS radial velocity time series as Barycentric Julian Dates (BJD), radial velocities, and associated uncertainties in m s$^{-1}$. 

\paragraph{Caption for Data S4.}
\textbf{MINERVA-Australis radial velocity time series.}
Machine readable file of MINERVA-Australis radial velocity time series as Barycentric Julian Dates (BJD), radial velocities, and associated uncertainties in m s$^{-1}$. 

\paragraph{Caption for Data S5.}
\textbf{PFS radial velocity time series.}
Machine readable file of PFS radial velocity time series as Barycentric Julian Dates (BJD), radial velocities, and associated uncertainties in m s$^{-1}$. 



\end{document}